\documentclass[12pt,preprint]{aastex}

\newcommand{\Dpm}{{_{_{ +}}\!\!{\cal D}\!\!\!_{_{ -}}}}

\newcommand{\arcsinh}{{\rm arcsinh}}
\newcommand{\cotanh}{{\rm cotanh}}

\received{}
\revised{}
\accepted{}
\ccc{}
\cpright{}{}

\slugcomment{To appear in ApJ {\bf 582}, No. 2 (2003)}

\shorttitle{MAGNETIC PINCHING OF THE HFT: I.}
\shortauthors{TITOV, GALSGAARD \& NEUKIRCH}

\begin{document}

\title{Magnetic Pinching of Hyperbolic Flux Tubes:\\
      I. Basic Estimations}

\author{V.S. Titov }
\affil{Theoretische Physik IV, Ruhr-Universit\"{a}t Bochum, 44780 
Bochum}
\email{st@tp4.ruhr-uni-bochum.de}

\and

\author{K. Galsgaard and T. Neukirch} 

\affil{School of Mathematics and Statistics,
University of St Andrews, St Andrews 
KY16 9SS, Scotland}

\begin{abstract}
  The concept of hyperbolic flux tubes (HFTs) is a generalization of the 
concept of separator field lines for coronal magnetic fields with 
a trivial magnetic topology. An effective mechanism of a current layer 
formation in HFTs is proposed.
  This mechanism is called magnetic pinching and it is caused by 
large-scale shearing motions applied to the photospheric feet of  
HFTs in a way as if trying to twist the HFT\@.
  It is shown that in the middle of an HFT such motions produce a 
hyperbolic flow that causes an exponentially fast growth of the current
density in a thin force-free current layer.
  The magnetic energy associated with the current layer that is built up
over a few hours is sufficient for a large flare.
  Other implications of HFT pinching for solar flares are discussed as well.
\end{abstract}


\section{INTRODUCTION}

Over the last decades it has become clear from both observational and
theoretical points of view that current sheet formation
in the corona is one of the key processes
for solar physics \citep{prk94,prfrb00}.
  In particular, it provides a temporary deposit around the current sheets 
of free magnetic
energy for solar flares and creates favorable conditions
for subsequent rapid conversion of this energy into other forms.
  The corresponding local growth of current density in
the current sheet formation process may stimulate the onset of 
plasma instabilities and the
development of anomalous resistivity.
  Turbulent dissipation of the current layer due to
such instabilities is thought to be a reason of thermal and
supra-thermal processes in solar flares \citep{som92}.

Typically, the coronal magnetic field is frozen into plasma and its
Maxwellian stresses are large enough to dominate over other
forces in active regions.
  This means that in such plasmo-magnetic configurations the magnetic
pressure and tension approximately balance each other.
  Also the transit time of perturbations through the corona is much
less than the characteristic time of photospheric motions.
  Therefore, the configurations have to evolve through a 
sequence of nearly force-free equilibria in response to the time variation
of the photospheric boundary conditions.
  Due to the frozen-in condition the topological structure of the 
magnetic field 
in the corona is not changed by horizontal photospheric motions.
  However, it may be changed by a vertical injection of a new magnetic 
flux through the photosphere.
  In the generic case this may lead to the appearance
of magnetic null points in the corona.
  The magnetic forces in the vicinity of the nulls are
too weak to withstand to large variations of the ambient magnetic
stress.
 Therefore, the corresponding neighborhoods of the nulls generally 
collapse in evolving fields, producing topologically accessible
current singularities \citep{prtit96}.

Similar processes have to occur in the vicinity of the so-called bald
patches \citep{titea93}, which are those segments of photospheric
polarity inversion lines (IL), where the field lines touch the
photosphere \citep{shf86}.
 In comparison with the case of the nulls the physics of the formation
of the current singularities is a bit different here:
 the current sheets are formed due to an ``attachment'' of the touching
field lines to the very heavy photospheric material \citep{low87, alam89, 
vea91}. 

Thus, the presence of the topological features such as null
points and bald patches is a {\it condition\/} for the formation  of current
singularities in coronal configurations.
 An analysis of magnetic field structures in solar flares shows,
however, that not all the observed event can be explained by
the presence of such topological features 
and that therefore this class of features has to be extended \citep{dem97}.
 A possible extension can be found by using the concept of field
line connectivity.
 Indeed, in the framework of this approach
the nulls and bald patches, as well as the separatrix
field lines emanating from them, can be detected by discontinuous jumps
of the field line connectivity.
 The desired extension then can be found 
by weakening the condition that the jump in connectivity is discontinuous
to requiring only that the field line connectivity exhibits
a ``large'' spatial variation 
\citep{shf86, lngcstr94}.
 The flux tubes which exhibit such a behavior are called
quasi-separatrix layers (QSLs) \citep{prdem95}.
 It should be emphasized that QSLs are geometrical objects rather than
topological ones, since they can be removed by suitable continuous
deformations of the magnetic field \citep{tithrn02}.
 The genuine separatrix lines and surfaces are 
degenerate or limiting cases of QSLs \citep{titea02}.

 For determining QSLs in a given magnetic field a measure of
magnetic connectivity is required. 
 The proper measure is {\it the degree of squashing\/} of elemental flux
tubes, which connect opposite photospheric polarities and have
infinitesimal cross-sections \citep{titea99}. 
 A QSL is then defined as a flux tube with abnormally large 
values of the squashing degree.
 In application to quadrupole magnetic configurations formed by two 
bipolar groups of sunspots this criterion reveals a special geometrical 
feature called {\it hyperbolic flux tube\/} (HFT) \citep{titea02}.
HFTs can be understood as a combination of two intersecting QSLs.
 In the limiting case, where the sunspot flux is concentrated in point-like 
sources, the HFT collapses into two separatrix surfaces intersecting 
along a separator field line.
 The separator is a field line connecting two null points which appear
in this limiting case and it is a favorable site for current
sheet formation caused by displacements of the sunspot positions
 \citep{swt69, gs88, lf90,lnc02}.
 Therefore it is natural to expect that HFTs are also preferred
sites for current sheet formation.  
 Here we will call such a process of current sheet formation in an HFT 
{\it  magnetic pinching\/}, by
analogy with a similar process studied in axisymmetric 
laboratory plasmas (the corresponding similarity will be further clarified
below).

It is the purpose of this series of papers 
to understand and quantify the basic properties of the process of
current sheet formation in HFTs. In the present paper we present basic
theoretical calculations and estimates relevant to this problem in the
following way.
 In \S~\ref{s:form} the problem and an idea for its solution are formulated.
 In \S~\ref{s:S2} a simplified quadrupole configuration used to model
an HFT is described. 
 In \S~\ref{s:km} a kinematic model of the HFT pinching process is
developed. 
 In \S~\ref{s:ffp} this model is improved by incorporating an
approximate form of magnetic force balance.
 In \S~\ref{s:impl} the implications of the obtained results for solar
flares are discussed.
 Section~\ref{s:cs} presents the conclusions of this work. 
 The results of dynamic and quasi-static simulations of the HFT
pinching process will be described in forthcoming papers of this series.

\section{FORMATION OF A CURRENT LAYER IN A QUADRUPOLE CONFIGURATION}
\label{s:form}

It is clear that the pinching of an HFT has to be due to some 
photospheric boundary conditions, which in the simplest form imply 
certain footpoint displacements of the field lines on the photosphere.
 \citet{prdem95} proposed that all types of 
displacements across QSLs must lead to the formation of a current
layer. 
 However, there are both theoretical arguments \citep{intit97} and numerical
simulations \citep{glsgd00} which
indicate that this is not the case -- the footpoint displacements
must be of a special type to cause the formation of current
layers. As HFTs consist of intersecting QSLs it is to be expected that
this result also has some importance for the pinching of HFTs.

To gain a better understanding of the problem of finding
those footpoint displacements which give rise to the pinching of HFTs, 
we first briefly discuss a particular aspect of the
phenomenology of large flares. 
 It was discovered a long time ago \citep{mrtr66, morsev68} that such
flares often occur in the vicinity of S-shaped ILs. 
On the basis of a potential field model using point sources
 \citet{gs88} proposed that this fact actually implies
the appearance of a current sheet forming along a
separator field line. The subsequent instability of this current sheet
results in a large flare.
In a quadrupole configuration this process is driven by 
suitable vortex motions near the IL\@.
As we will show below this idea also works, under suitable modifications,
if one replaces the 
separator field line by an HFT.

It is important to notice that in a realistic configuration with 
distributed photospheric flux the appearance of a separator in the
corona is quite problematic, because it requires the simultaneous
appearance of two magnetic nulls in the corona.
For the presence of an HFT, however, much less restrictive conditions
apply, so that it is a generic geometrical feature for quadrupole
configurations \citep{titea02}.

\begin{figure}
\epsscale{0.6}
\plotone{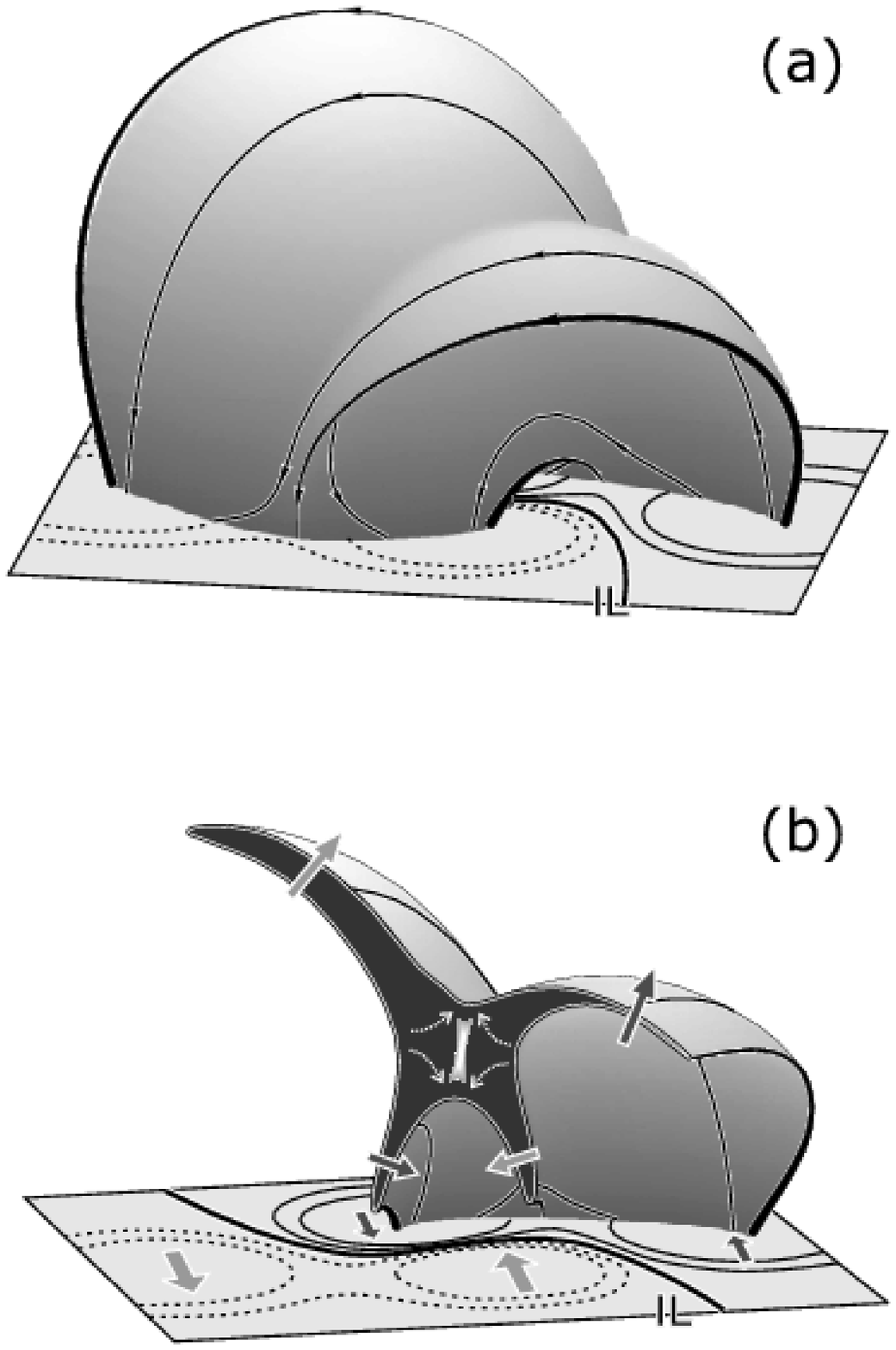}\\
\caption{Hyperbolic flux tube (a) and half of it (b) in a quadrupole
configuration.
 The varying thickness of the field lines at the ``ribs'' of the HFT
depicts the corresponding variation of the field strength or their
``rigidity''.
 The large arrows on (b) show the directions of plasma
flows which lead to the current layer formation in the middle of the
HFT\@.
 The flows in the corona are due to the appropriate sunspot motions and
elastic properties of the HFT\@.
 \label{f:Q100xxx} }
\end{figure}

The structure of a typical HFT is shown in Figure~\ref{f:Q100xxx}a.
Along an HFT its 
cross-section typically changes as shown in the
following schematic sketch
   $$\hbox{}
\begin{picture}(125,8)(0,0)
  \multiput(0,0)(-0.3,0){6}{\line(1,1){10}}
      \put(14,2){$\rightarrow$}
  \multiput(28.15,2)(-0.28,0){5}{$\times$}
  \multiput(28,0)(-0.3,0){6}{\line(1,1){10}}
      \put(42,2){$\rightarrow$}
  \multiput(56,0)(-0.3,0){6}{\line(1,1){10}}
  \multiput(66,0)(-0.3,0){6}{\line(-1,1){10}}
      \put(71,2){$\rightarrow$}
  \multiput(95,0)(-0.3,0){6}{\line(-1,1){10}}
  \multiput(85.5,2)(-0.28,0){5}{$\times$}
      \put(100,2){$\rightarrow$}  
  \multiput(124,0)(-0.3,0){6}{\line(-1,1){10}}
\end{picture}\hbox{}.
   $$
The intersections of the HFT with the photosphere
have a crescent-like shape with the tips of the crescents located inside the
neighbouring sunspots of the same polarity.
Thus, the photospheric field strength has minima in the middle of the
crescents and increases significantly towards its tips.
 One footpoint of any field line inside the HFT is always close to one
of the tips, while the other footpoint is in the middle of the
opposite polarity crescent.   
 In other words, each field line inside the HFT connects regions of 
strong and weak photospheric field of opposite polarity.
 Along each field line the field strength and the Maxwellian stresses 
increase or decrease  
monotonically, depending on the direction of the field line from weak to
strong field or vice versa.
 Towards the strong field region field lines will be 
relatively rigid becoming more flexible towards the weaker field region.
 In Figure~\ref{f:Q100xxx}a this property is depicted by the
corresponding variation of the field line thickness.

Such a variation of field line rigidity within the HFT 
predetermines the structure of
the three-dimensional deformation caused by displacements
of the foot points.
 One can expect that the farther the field line remains inflexible from 
its rigid end, the farther the photospheric displacement from this end 
propagates into the corona.
 This implies, in particular, that a shearing motion across the
crescent-like foot of the HFT easily propagates to its middle
part in the corona. 
 The same is true, of course, for the second foot.
 Since the cross-section of the HFT in its middle has an X-like shape, 
the shears propagating from the feet to the middle will superpose 
there perpendicular to each other.
If the foot point motions at the feet are
applying twist to the HFT, then their superposition will cause a hyperbolic
flow pattern in the centre of the HFT (Figure~\ref{f:Q100xxx}b). 
Photospheric shearing motions are caused quite naturally by sunspot
displacements perpendicular to the HFT feet and
the twisting type of shearing motion
corresponds exactly to the above mentioned S-like deformation of the
IL by photospheric flows. 
Thus, the hyperbolic flow pattern in the middle of the HFT will persist as
long as such sunspot displacements continue.
As we will show this hyperbolic flow pattern causes an exponentially fast
deformation of the HFT into a sheet-like structure (see below).
Therefore, if the total length of the foot point displacements 
is comparable with the size of the active
region one can expect a very large pinching of the middle part of the HFT
due to the hyperbolic flow pattern.

We conclude that a combination of twisting foot point displacements 
perpendicular to the foot points of the HFT is 
sufficient for the formation of a strong current layer in the HFT\@.
 In this case the velocity field and transverse magnetic field component
inside the HFT will have a similar hyperbolic structure but are
inclined by 45$^{\circ}$ with respect to each other. 
 In the theory of two-dimensional (2D) reconnection these hyperbolic
structures of the velocity and magnetic field are the main
prerequisites for the so-called magnetic collapse 
\citep{prfrb00} or the dynamic formation of a current sheet \citep{syr81}.
The idea that current layers can form in three-dimensional (3D) magnetic
configurations under the action of hyperbolic flows was also
considered earlier.
 \citet{glsnrd96} imposed two shearing motions of a fixed duration,
one after another in mutually perpendicular directions, onto an initially
uniform magnetic field.
 This forced the magnetic field lines to twist around each other, so
that the magnetic tension stress appearing in this process sets up a
stagnation flow halfway between the driving boundaries.
 The action of this flow during the corresponding time interval led to
the formation of a strong current layer.
 To form such current layers \citet{cowea97} assumed the presence of
stagnation flows on the photosphere driving the system.
 However, sustaining these flows for sufficiently
long time seems to be rather problematic.
 On the contrary, the twisting shears discussed above appear as a
result of the appropriate displacements of the sunspots perpendicular
to the HFT feet and can be part of the natural photospheric
motions having a relatively long lifetime.
 Due to these arguments the hyperbolic flows generated inside an HFT
are generic features of HFTs just as HFTs themselves are for
quadrupolar magnetic field configurations.

\section{AN HFT IN A SIMPLIFIED QUADRUPOLE CONFIGURATION} \label{s:S2}

To facilitate the quantitative investigation of the problem, we
simplify the quadrupole configuration discussed above in the 
same way as it is usually done to model magnetic flux braiding
\citep{prk72}. 
We assume that the coronal plasma and field are contained between 
two planes, $z=-L$ and $z=L$, representing positive and
negative photospheric polarities, respectively. 
Let the initial magnetic field be current-free and produced by four
fictive magnetic charges placed outside of the model volume.
The sources are all equal in strength but the four sources
fall in two pairs which differ in polarity.
The negative and positive pairs are located at the points ${\bf
r}_{1,3}=(\pm l,0,L+d)$ and ${\bf r}_{2,4}=(0, \pm l, - L-d)$, respectively.
Their strength is chosen to be $B_{\rm s}\, d^2$, so that the normal  
component of the field at the photosphere would have extrema $\approx\pm 
B_{\rm s}$ when their depths are small ($d \ll l,L$).
  Then the resulting initial magnetic field is given by
 \begin{eqnarray}
  {\bf B}_{0}({\bf r}) =
 B_{\rm s}\, d^2 \sum_{n=1}^{4} (-1)^{n} {({\bf r} - {\bf r}_{n}) 
 \over |{\bf r} - {\bf r}_{n}|^{3}}.
        \label{B0}
 \end{eqnarray}

\begin{figure}
\epsscale{0.4}
\plotone{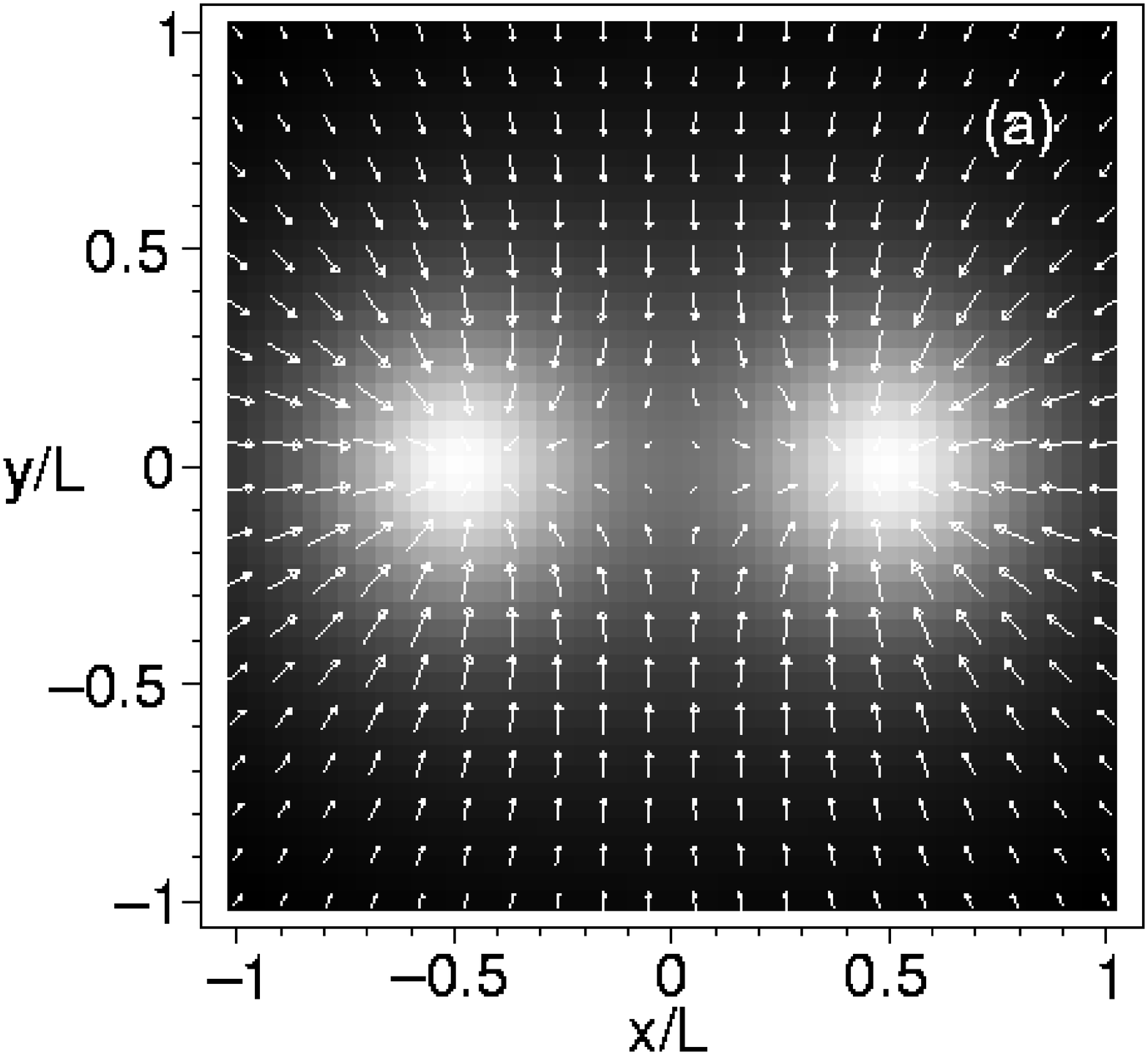}\\
\plotone{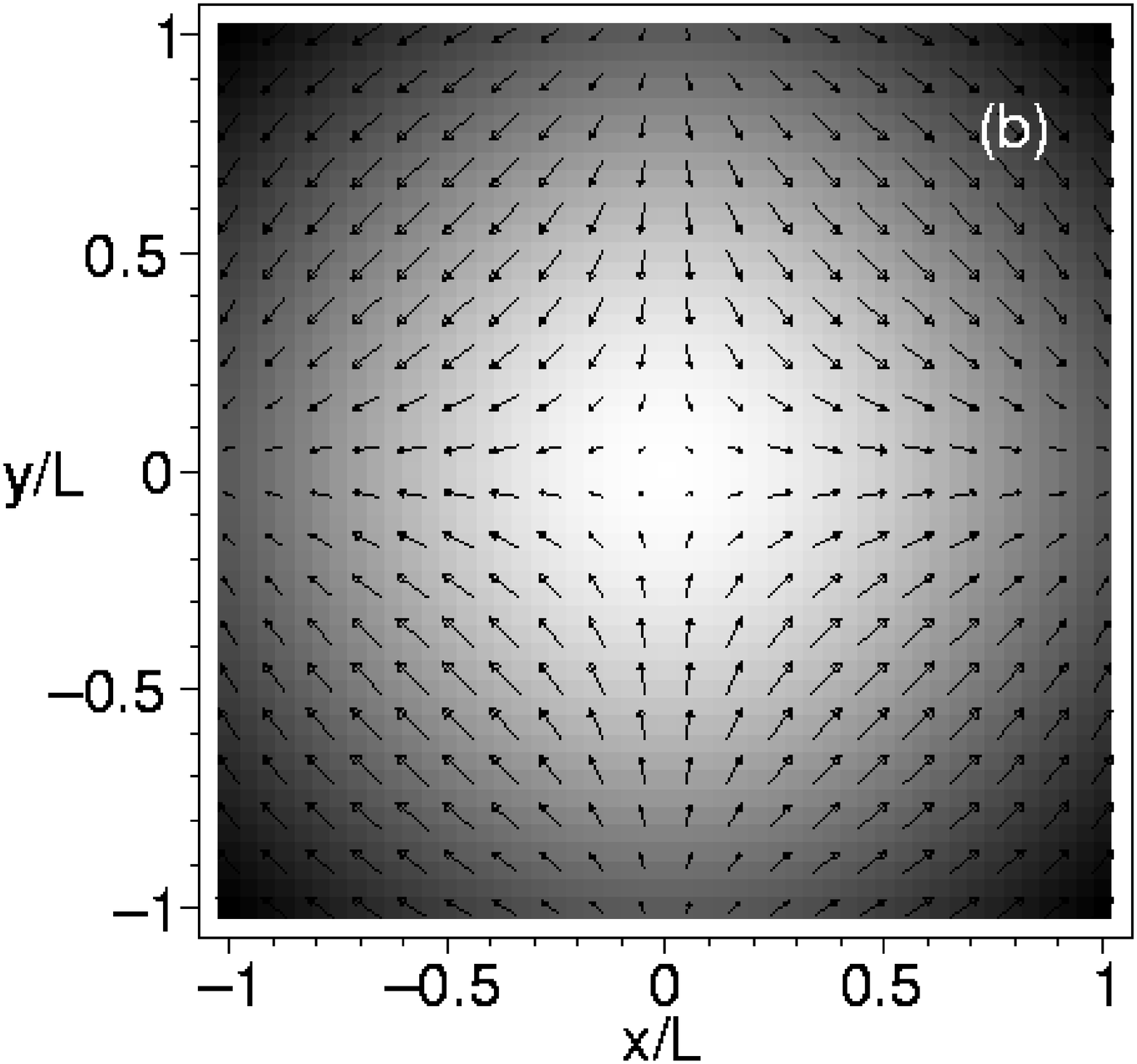}\\
\plotone{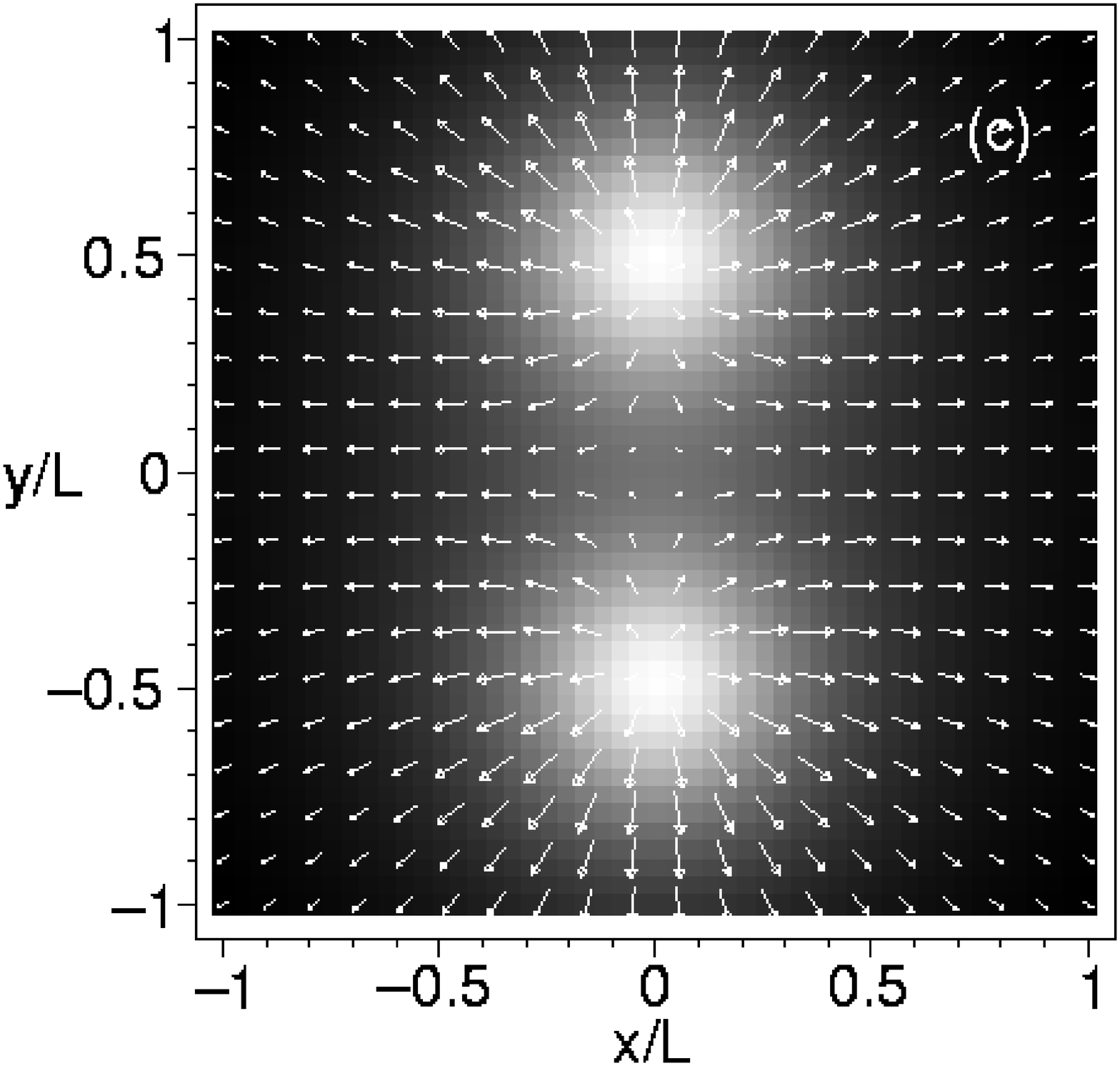}
\caption{The effective magnetograms of the model configuration at
$l/L=0.5$, $d/L=0.4$ in the planes  $z=L$ (a),
$z=0$ (b) and $z=-L$
(c).  The distribution of the $B_{0z}$-component is shown in grey
half-tones, while vectors represent the transverse magnetic field. The
intensity of the grey half-tones and the length of the vectors are
normalized by the corresponding maximum values in the planes.
 \label{f:emgs} }
\end{figure}

This field has a trivial topological structure in the corona, since
there are no nulls in the coronal volume $|z|<L$.
 However, owing to the presence of two flux concentrations in each
photospheric plane (cf. Figure~\ref{f:emgs}a and~\ref{f:emgs}c) the
geometry of the field is not simple in general.
 Such concentrations are smoothed out by a large expansion of the flux
tubes between the planes to yield a rather homogeneous
$B_{0z}$-distribution in the midplane $z=0$ (Figure~\ref{f:emgs}b).
The pairs of concentrations have an orientation which is perpendicular
to each other and therefore the transverse field $(B_{0x},B_{0y})$ remains
inhomogeneous in the plane $z=0$ and has the typical structure of a
hyperbolic magnetic X-point (Figure~\ref{f:emgs}b).

This allows us to approximate the magnetic field
near this plane by
 \begin{eqnarray}
  {\bf B}_{0} \simeq (hx,-hy, B_{\|}),
  \label{B0a}
 \end{eqnarray}
where the gradient of the transverse field, $h$, and the longitudinal
field, $B_{\|}$, are easily derived from~(\ref{B0}) to yield
 \begin{eqnarray}
   h & = & {6 (B_{\rm s}/L) \bar{d}^{2} \bar{l}^{2} \over  \left[
       \bar{l}^2 +  (1+\bar{d})^2
     \right]^{5/2}},
   \label{h} \\
  B_{\|} & = & {4 B_{\rm s} \bar{d}^2 (1+\bar{d})
    \over  \left[ \bar{l}^2 + (1+\bar{d})^2 \right]^{3/2}}.
  \label{Bpar}
 \end{eqnarray}
The bar over the length scales denotes their normalization to~$L$.
 This approximation is fairly accurate for
$r,|z| \lesssim l$ and will be useful for our further analysis of the
problem.

This configuration is constructed in such a way that it
represents a simplified or
``straightened'' version of the HFT considered in the previous section.
If we use the squashing degree of elemental flux tubes in QSLs as
proposed by 
\citet{titea99},
 \begin{eqnarray}
  Q={ \left( {\partial X_{-}\over \partial x_{+}}\right)^2
     +\left( {\partial X_{-}\over \partial y_{+}}\right)^2
     + \left( {\partial Y_{-}\over \partial x_{+}}\right)^2
     +\left( {\partial Y_{-}\over \partial y_{+}}\right)^2
 \over
 \left|
  {\partial X_{-}\over \partial x_{+}}
  {\partial Y_{-}\over \partial y_{+}}
 -{\partial X_{-}\over \partial y_{+}}
  {\partial Y_{-}\over \partial x_{+}}
 \right| },
        \label{Q}
 \end{eqnarray}
where the functions $X_{-}(x_{+},y_{+})$ and
$Y_{-}(x_{+},y_{+})$ represent the connection
between the field line footpoints ($(x_{+},y_{+})$ and $(x_{-},y_{-})$),
the surface of the HFT is defined by the condition
$Q=\mbox{const}$ with the constant $\gg2$.

\begin{figure}
\epsscale{0.6}
\plotone{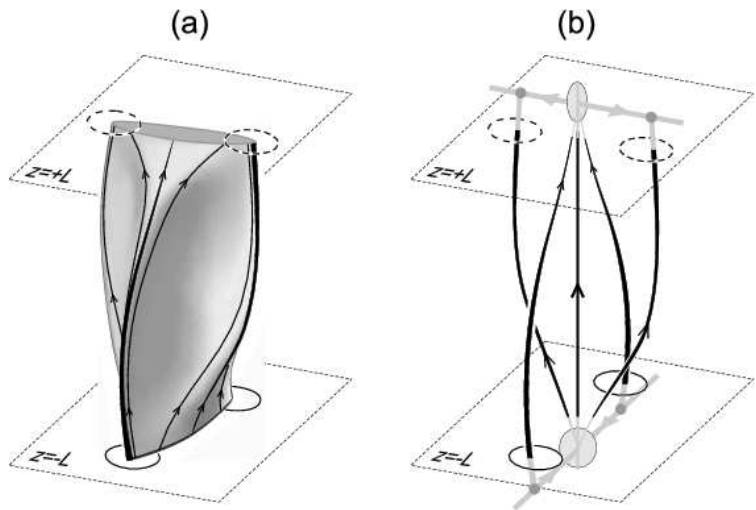}
\caption{The HFT in a simplified configuration (a) and the
topological structure of the field including
the fictive magnetic charges and nulls which are located outside the
model volume.
 The circles in the planes $z=\pm L$ represent the sunspots, while the grey
circles (b) show the separatrix fan planes at the nulls, from where
the separatrix spine lines emanate towards the charges.
 The varying thickness of the field lines depicts the corresponding
variation of the field strength or their ``rigidity''.
 \label{f:hft_str} }
\end{figure}

The value $Q$ is invariant to interchanging the 
$+$ and $-$ signs in (\ref{Q}). 
A general way to compute $Q$ numerically is to calculate field lines
starting at neighboring footpoints and follow them to their other
footpoints. Then the respective
derivatives in~(\ref{Q}) can be calculated.
 A sketch of the typical shape of the surface enclosing the HFT is
presented in Figure~\ref{f:hft_str}a, which shows that the
intersection of the HFT with the photospheric planes is very narrow in
one direction and elongated in the other, connecting the two
flux concentrations.
 The surfaces calculated using other values of $Q$ have similar shapes 
and are nested
inside each other with the corresponding value of $Q$ growing 
as the surfaces approach the $z$-axis.
 The magnitudes of the magnetic field and the Maxwellian stresses 
 in this ``straightened'' HFT 
are distributed in a similar way as in the more realistic curved HFT version
described above.

\begin{figure}
\epsscale{0.6}
\plotone{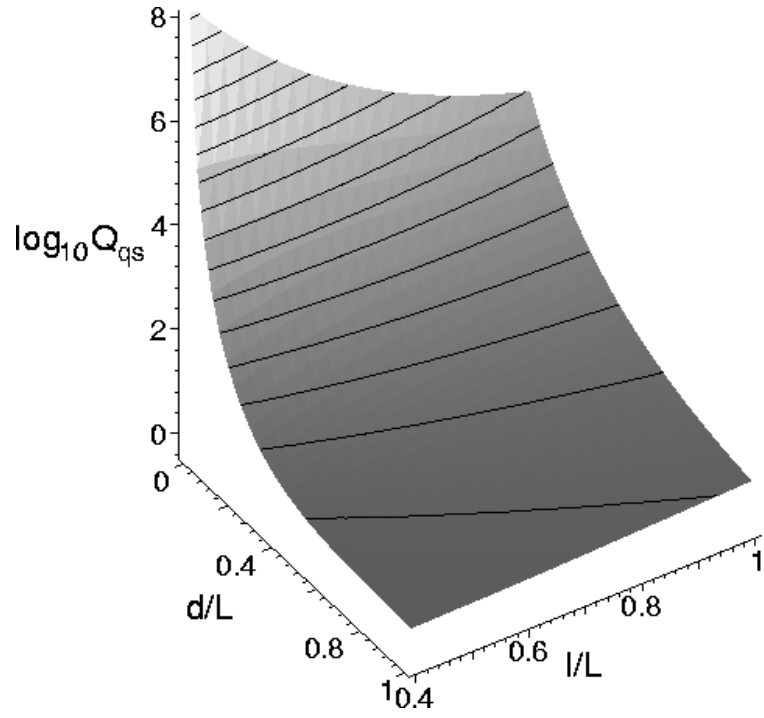}
\caption{The logarithm of the squashing degree  $Q_{\rm qs}$ at the
quasi-separator as a function of the depth $d$ and half-distance $l$
between the sources.   The length scales are normalized to the
half-length $L$ of the HFT and the equidistant contours are plotted with an
increment of $0.5$.
 \label{f:lgQz} }
\end{figure}

To estimate the range of parameters for which the HFT is
present, we calculate the maximum of $Q$ within the configuration.
To do this we start by noticing that if we  extend our model magnetic field
(\ref{B0})
to the region outside the model volume, then the $z$-axis coincides with
a separator field line for this extended field because it has two null
points on the $z$-axis outside the model volume (Figure~\ref{f:hft_str}b).
Therefore, by construction, the maximum of $Q$ inside the model volume is 
attained on the field line 
coinciding with the $z$-axis.
In the following, we call this field line the {\it quasi-separator\/} and 
denote the corresponding squashing degree by $Q_{\rm qs}$.
Details of the derivation of $Q_{\rm qs}$ are given in Appendix A, with the
result being
 \begin{eqnarray}
  Q_{\rm qs} = 2 \cosh 2\lambda,
        \label{Qqs}
 \end{eqnarray}
where
 \begin{eqnarray}
  \lambda = \int_{-L}^{+L} \left. \left( {\partial B_{0x} \over 
\partial  x} \bigg/ 
B_{0z} \right)\right|_{{\bf r}=(0,0,z)} \, {\rm d}z.
        \label{lmd}
 \end{eqnarray}
 Thus, $Q_{\rm qs}$ grows exponentially with
$\lambda$, which in turn grows if the gradient of
the transverse field component increases and/or the longitudinal
component decreases on average along the quasi-separator.
 The explicit form of the integrand in (\ref{lmd}) is cumbersome but 
easily derivable from~(\ref{B0}).
 Since the integral itself is not expressible in terms of elementary
or special functions, we do not give
further details of the integrand and simply present the
corresponding numerical results of the
integration in Figure~\ref{f:lgQz}.
It can be seen that $Q_{\rm qs}$ strongly grows with a
growing concentration of
flux in the sunspots (i.e. decreasing $d/L$) and if the distance of
the sunspots of the same polarity decreases (i.e. decreasing $l/L$).
 In other words, the stronger and the closer the sunspots in our
configuration are, the larger will be the squashing degree $Q_{\rm qs}$
 at the quasi-separator.
Figure~\ref{f:lgQz} of $Q_{\rm qs}$ indicates 
that the HFT
should exist in this configuration for a relatively large range of
parameters of the model.

\section{KINEMATICS OF HFT PINCHING} 
\label{s:km}

For the time scale during which the sunspots are passing each other,
the flow velocity can be
assumed to be constant and uniform in their vicinity.
 Between two spots moving in opposite directions a
region of shearing flow must exist to provide a continuous velocity field.
 We restrict ourselves to the cases, where the spots
of the same polarity move perpendicularly to the HFT feet.
 Taking for simplicity the characteristic velocity $V_{\rm
s}$ to be the same for all spots, we define the photospheric
velocity field by
 \begin{eqnarray}
 {\bf v}_{+} = V_{\rm s} \tanh(y/l_{\rm sh})\, \hat{\bf x}
 \qquad\mbox{at\ } z=-L
 \label{vdwn}
 \end{eqnarray}  
and
 \begin{eqnarray}
 {\bf v}_{-} =\mp V_{\rm s} \tanh(x/l_{\rm sh})\, \hat{\bf y} 
\qquad\mbox{at\ } z=L,
 \label{vup}
 \end{eqnarray} 
where $l_{\rm sh}$ is a characteristic length scale of the shear, and
we have kept the possibility to change the sign of the second velocity
field to demonstrate the effects of favorable and unfavorable foot point
motions (see below).
 Assuming that the shearing motions approximately match the corresponding
motions of the spots at the locations where
$B_{z}^{2} \approx B_{{\rm s}}^{2}/2$, we can estimate $l_{\rm sh}$
at $d<l$ as
 \begin{eqnarray}
  l_{\rm sh} \approx l-0.6\,d.
  \label{lsh}
 \end{eqnarray}
Depending on the sign ($-$ or $+$) in (\ref{vup}), we either have a
``turning'' or a ``twisting'' pair of shearing motions, respectively,
which are applied to the feet of the HFT.  
It is crucial for our
understanding of the pinching process of the HFT to consider these 
two fundamental cases.  
More general motions of the spots can be obtained 
as linear combinations of these two.

As discussed above, inside the photospheric boundaries the
magnetic field strength grows from the middle
of the HFT feet towards their ends which are located at the magnetic
flux concentrations.
 Also field lines belonging to the HFT always connect regions of strong
and weak magnetic field.
 Since the rigidity of the field lines is proportional to the local field 
strength, the HFT field lines are rigid at one footpoint 
and flexible at the other. 
 These elastic properties of the HFT magnetic field 
strongly suggest that the effect of the 
photospheric shearing motion must propagate along the field lines 
from one boundary to the other
but with gradually vanishing amplitude.
We approximately model the resulting three-dimensional velocity field 
by extrapolating both (\ref{vdwn}) and (\ref{vup})
linearly into the volume and then superposing them with the
result
 \begin{eqnarray}
  {\bf v} ={ V_{\rm s} \over 2}
 \left[ \left(1-{z\over L}\right) \tanh\left({y\over l_{\rm sh}}
        \right)\, \hat{\bf x} 
    \mp \left(1+{z\over L}\right) \tanh\left({x\over l_{\rm sh}}
        \right)\, \hat{\bf y} \right].
        \label{vcor}
 \end{eqnarray}
This velocity field yields the simplest model for infinitesimal
deformations (${\bf 
  v}\, {\rm d}t$) of the HFT over an infinitesimal time interval ${\rm 
d}t$ in response to the ``turning'' or
``twisting'' pair of photospheric shearing motions ${\bf v}_{+}\, {\rm
  d}t$ and ${\bf v}_{-}\, {\rm d}t$.
 The sign ($-$ or $+$) of the second term of~(\ref{vcor}) corresponds to
``turning'' or ``twisting'' motion, respectively.
The velocity fields (\ref{vcor}) are not exact solutions of the MHD
equations, of course. 
 Nevertheless, they are qualitatively consistent with the elastic 
properties of the HFT and hence studying the effect of
these velocity fields provides useful 
insights into the physical mechanism of the HFT pinching process.

With this purpose we now investigate the effect of finite deformations of the
HFT by the velocity fields (\ref{vcor}).
Finite deformations are described by the vector 
function ${\bf r}(t,{\bf r}_{0})$ in
which ${\bf r}_{0} \equiv (x_{0}, y_{0}, z_{0})$ represents the
Lagrangian coordinates of plasma elements.
 This function ${\bf r}(t,{\bf r}_{0})$ is obtained from 
(\ref{vcor}) by integrating the equation
 \begin{eqnarray}
  {{\rm d} {\bf r} \over {\rm d} t} = {\bf v}
        \label{ef}
 \end{eqnarray}
with the initial condition ${\bf r}(0,{\bf r}_{0})={\bf r}_{0}$.
 Since the velocity fields (\ref{vcor}) have a
vanishing $z$-component, the corresponding 
deformations have only $x$- and $y$-components and the plasma
elements remain in the same plane $z=\mbox{const}$ in which they
were initially.
It is of particular interest to study these deformations in the
midplane $z=0$, where the transverse magnetic field has a structure typical 
for the neighborhood of an X-point.

\begin{figure}
\epsscale{0.8}
\plotone{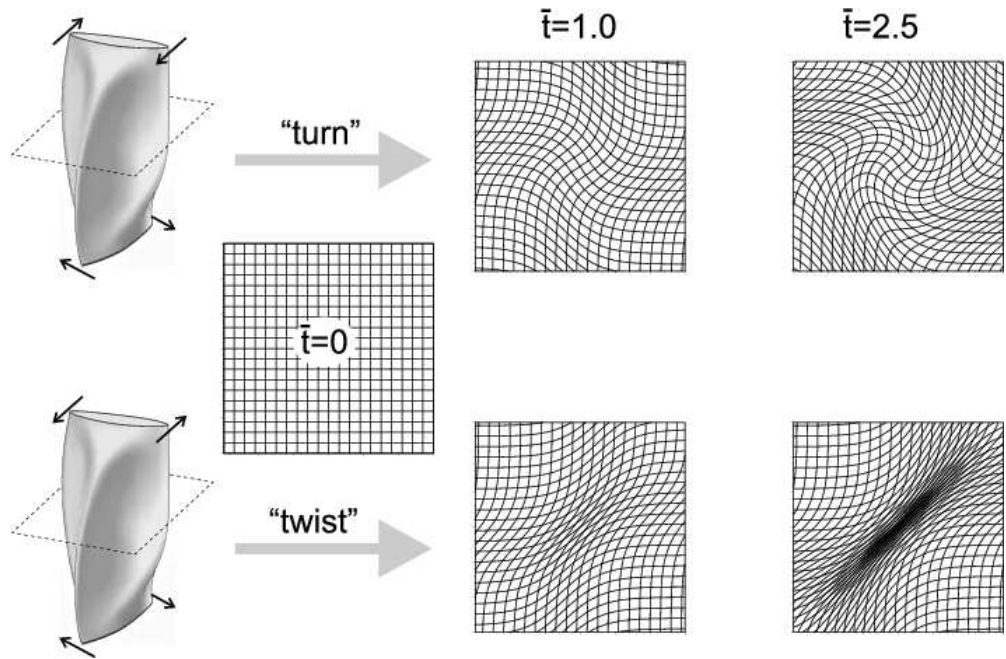}
\caption{Non-pinching (top) and pinching (bottom) deformations of 
the HFT in the (dash-outlined) plane $z=0$ due to
``turning'' and ``twisting'' shearing motions, respectively, applied
to the HFT feet.
 \label{f:dfms} }
\end{figure}

\begin{figure}
\epsscale{0.4}
\plotone{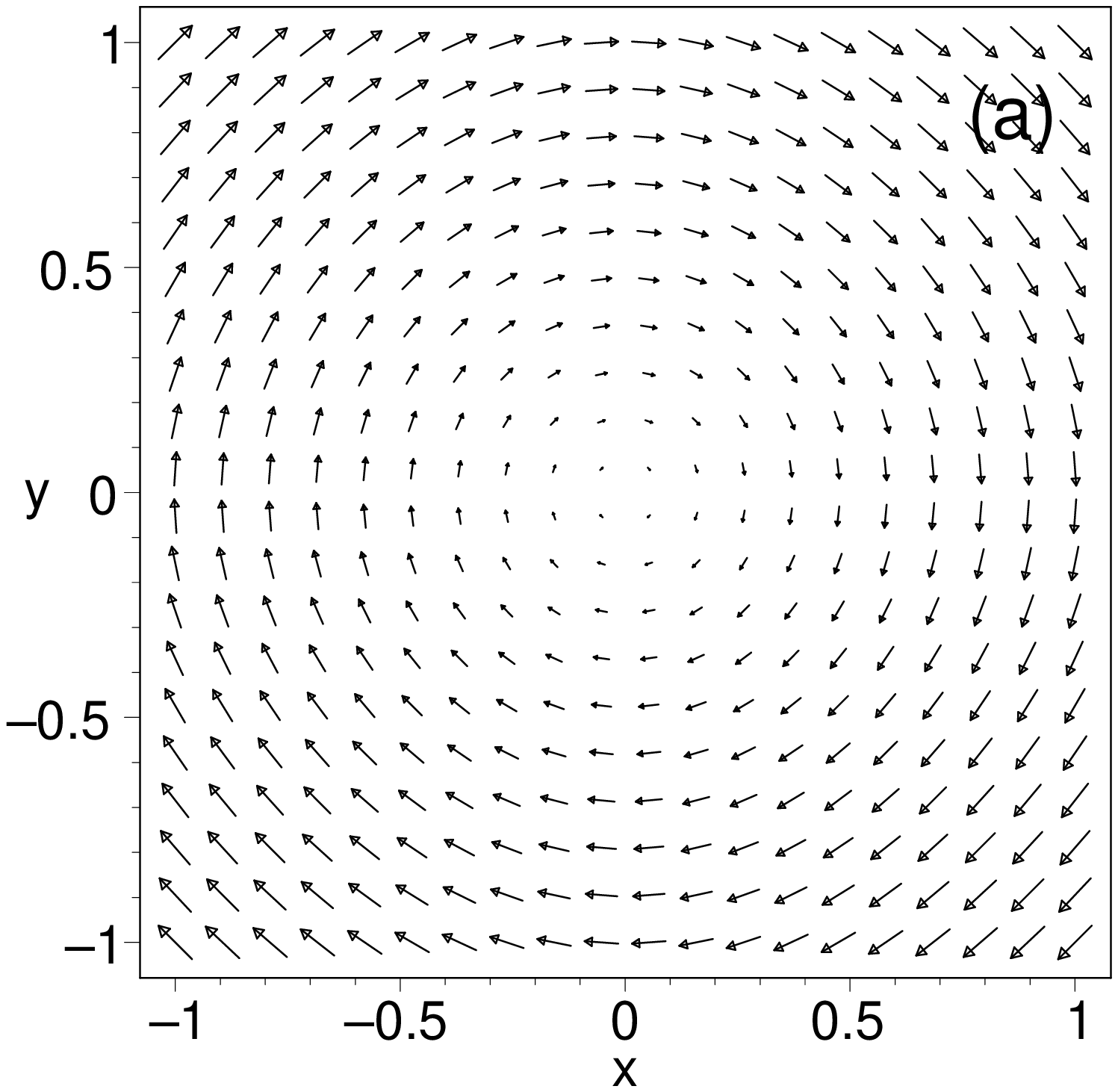}\\
\plotone{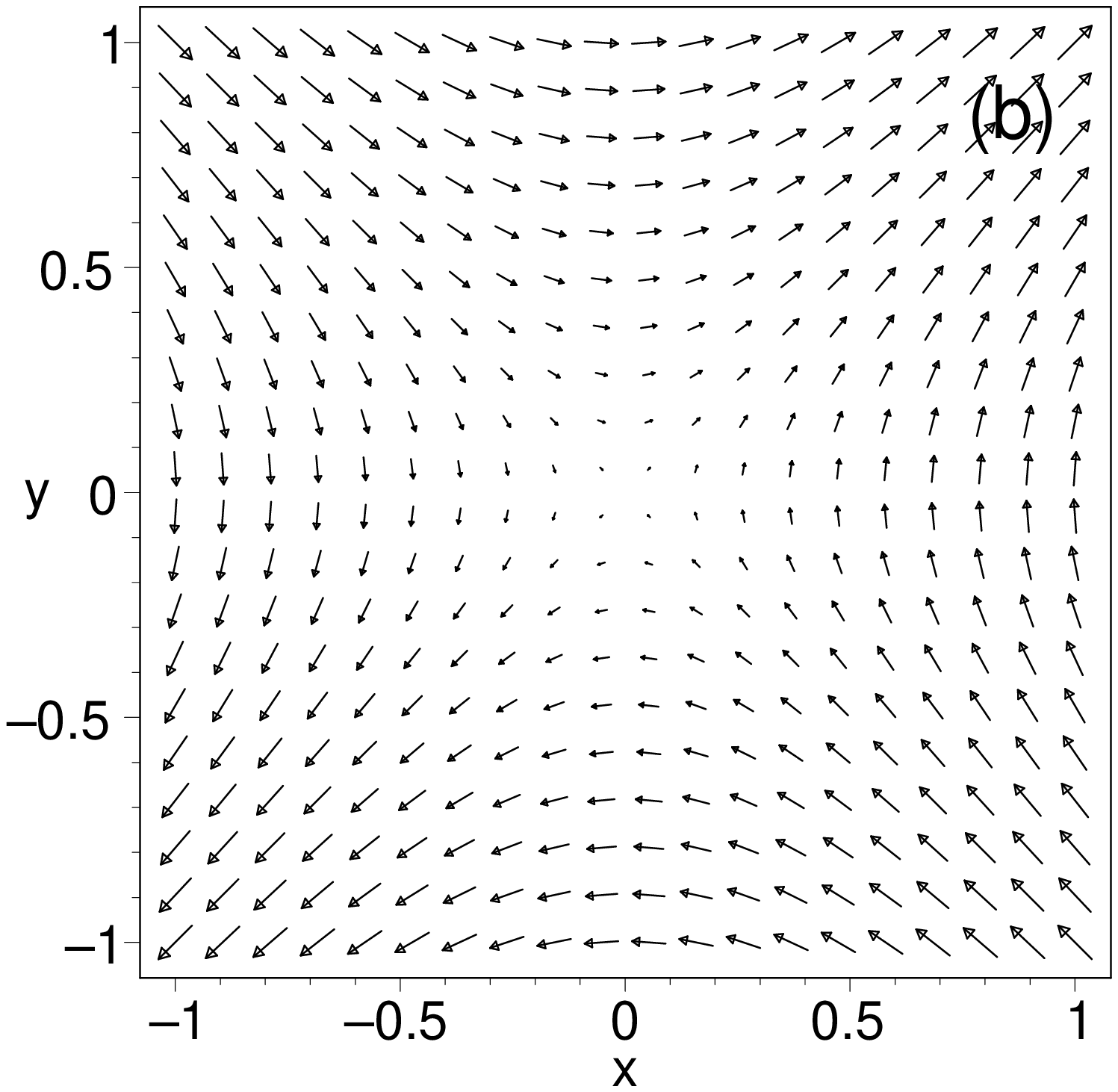}
\caption{Rotational (a) and hyperbolic (b) velocity fields in the
plane $z=0$ due to ``turning'' and ``twisting''  shearing motions,
respectively, applied to the HFT feet.
 \label{f:vfz0} }
\end{figure}

We have done this by computing the corresponding distortion of an 
initially uniform grid of plasma elements for two subsequent moments.
 The results are represented in Figure~\ref{f:dfms}, which shows that 
``turning" shearing motions cause only a relatively slight distortion
of the grid even if the sunspots move over distances of the order of~$L$.
 The middle part of the HFT in this case just turns as a whole by an
angle of $\sim t$ with only modest changes in the sizes of the grid cells.
 By contrast, the ``twisting"  shearing motions produce a much
stronger effect in the central parts of the grid
causing an extreme squashing of the grid cells in one direction.
 The reason for this difference becomes clear if one compares the
corresponding velocity fields in the plane~$z=0$.  
 In the first case the velocity field has the typical structure of 
rotational flows (see Figure~\ref{f:vfz0}a),
 while in the second case it shows the velocity field of a
stagnation point (Figure~\ref{f:vfz0}b).
Especially the hyperbolic structure of the stagnation type velocity field
causes an
exponentially fast pinching of the HFT into a thin layer.

To show this behavior more explicitly, we have solved Eq.~(\ref{ef})  
at $z=0$ analytically with the
result
 \begin{eqnarray}
  \bar{x}  =  {\rm arcsinh\: }[\cosh \bar{x}_{0}\, ( \cosh \bar{t} \, 
\tanh \bar{x}_{0}\, +&&
     \nonumber \\
                            \sinh \bar{t}\,  \tanh \bar{y}_{0} ) ], && 
        \label{Xz0}  \\
  \bar{y}   =  {\rm arcsinh\:}[\cosh \bar{y}_{0}\, ( \cosh \bar{t} \, 
\tanh \bar{y}_{0}\, + &&
   \nonumber \\
                             \sinh \bar{t}\,  \tanh \bar{x}_{0} ) ]. &&
        \label{Yz0}
 \end{eqnarray}
Here we have used
 \begin{eqnarray}
  \bar{t}= {V_{\rm s} t\over 2\,l_{\rm sh} },
  \label{tn}
 \end{eqnarray}
while $\bar{x}=x/l_{\rm sh}$,  $\bar{x}_{0}=x_{0}/l_{\rm sh}$ and
similarly for the other spatial variables.
The lengthy derivation of this solution is omitted here, but it can
be verified in a straightforward manner by direct substitution into
(\ref{ef}). 

Mathematically the obtained deformations (\ref{Xz0}) and (\ref{Yz0}) form a
one-parameter group of diffeomorphisms, with the parameter~$t$.
 So if ${\mathcal F}_{t_{1}}$ and ${\mathcal F}_{t_{2}}$ are 
elements of this group corresponding to times $t_{1}$ and $t_{2}$,
respectively, their composition is ${\mathcal F}_{t_{2}} 
\circ {\mathcal F}_{t_{1}} = {\mathcal F}_{t_{1}+t_{2}}$.
 In particular, ${\mathcal F}_{-t} \circ {\mathcal F}_{t} = {\mathcal
F}_{0} \equiv {\mathcal Id}$, which means that one can reverse the 
deformations 
by changing the variables in (\ref{Xz0})--(\ref{Yz0}) as
$\bar{t} \rightarrow -\bar{t}$,
$\bar{x} \leftrightarrow \bar{x}_{0}$ and $\bar{y} \leftrightarrow
\bar{y}_{0}$ to get
 \begin{eqnarray}
  \bar{x}_{0}  =  {\rm arcsinh\: }[\cosh \bar{x}\, ( \cosh \bar{t} \, 
\tanh \bar{x}\, -&&
     \nonumber \\
                            \sinh \bar{t}\,  \tanh \bar{y} ) ], && 
        \label{X0z0}  \\
  \bar{y}_{0}   =  {\rm arcsinh\:}[\cosh \bar{y}\, ( \cosh \bar{t} \, 
\tanh \bar{y}\, - &&
   \nonumber \\
                             \sinh \bar{t}\,  \tanh \bar{x} ) ]. &&
        \label{Y0z0}
 \end{eqnarray}

\begin{figure}
\epsscale{0.4}
\plotone{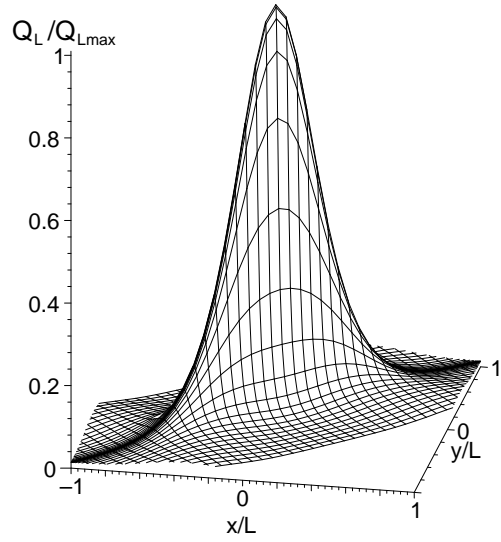}
\caption{The degree of squashing $Q_{\rm L}$ of Lagrangian plasma
elements in the plane $z=0$ at $\bar{t}=2.5$ (other parameters:
$d/L=0.4$ and $l/L=0.5$) for the deformation caused by the ``twisting''
photospheric shearing motion.
 \label{f:QL} }
\end{figure}

By construction, the velocity field (\ref{vcor}) and  the corresponding
finite deformations (\ref{Xz0})--(\ref{Y0z0}) are incompressible.
 So the plasma elements are distorted by these deformations mainly
due to compression in one direction and stretching in the other.
We can measure such a distortion by again using the degree of
squashing (\ref{Q}) defined earlier, 
but using Lagrangian mappings (\ref{Xz0})--(\ref{Yz0}) or 
(\ref{X0z0})--(\ref{Y0z0}) instead of the 
field line mapping.
Due to the incompressibility of the Lagrangian deformations, 
the corresponding degree of squashing
$Q_{\rm L}$ has simpler form than $Q$, given by
 \begin{eqnarray}
  Q_{\rm L} = 
  \left( {\partial x_{0} \over \partial x } \right)^{2}
 +\left( {\partial x_{0} \over \partial y } \right)^{2}
 +\left( {\partial y_{0} \over \partial x } \right)^{2}
 +\left( {\partial y_{0} \over \partial y } \right)^{2}.
 \label{QL}
 \end{eqnarray}
A plot of the distribution of $Q_{\rm L}$ in the $(x,y)$-plane at
$z=0$ clearly demonstrates that the squashing of plasma elements is
concentrated in a layer-like region inclined to the $x$-axis by
$45^{\circ}$ (Figure~\ref{f:QL}).
 We define the thickness $\delta$ and the width $\Delta$ of this layer 
as the minimum and maximum extension of the contour $Q_{\rm
  L}(x,y)=Q_{\rm L\,max}/2$ in the $(x,y)$-plane, respectively. 
Here
 \begin{eqnarray}
   Q_{\rm L\, max} = 2 + 4\, \sinh^{2}\bar{t}
   \label{QLm}
 \end{eqnarray}
is the maximum of $Q_{\rm L}$ which is located at ${\bf r}=0$. 
 For sufficiently large times $\bar{t} \gtrsim 1$ we obtain
 \begin{eqnarray}
  \delta & \simeq & 2\sqrt{2}\, e^{-\bar{t}}  l_{\rm sh},  
  \label{dlt}\\
   \Delta & \simeq &  2.5\, l_{\rm sh}.
   \label{Dlt}
 \end{eqnarray}
 Thus, under a ``twisting" shearing motion applied to the 
HFT feet a layer-like structure forms in the middle of the 
HFT. The thickness of this layer decreases exponentially fast in time,
while its width remains approximately constant.
 
As we will show in \S~\ref{s:ffp} this effect is actually enhanced
if magnetic forces are taken into account.
 In many respects this process resembles the pinching of a magnetic rope
in laboratory plasmas, which has been realized in 2D theory of current
sheets long time ago \citep{syr81}.
Our model generalizes this process to 3D and for this reason we gave it  
the name magnetic pinching.

\begin{figure}
\epsscale{0.4}
\plotone{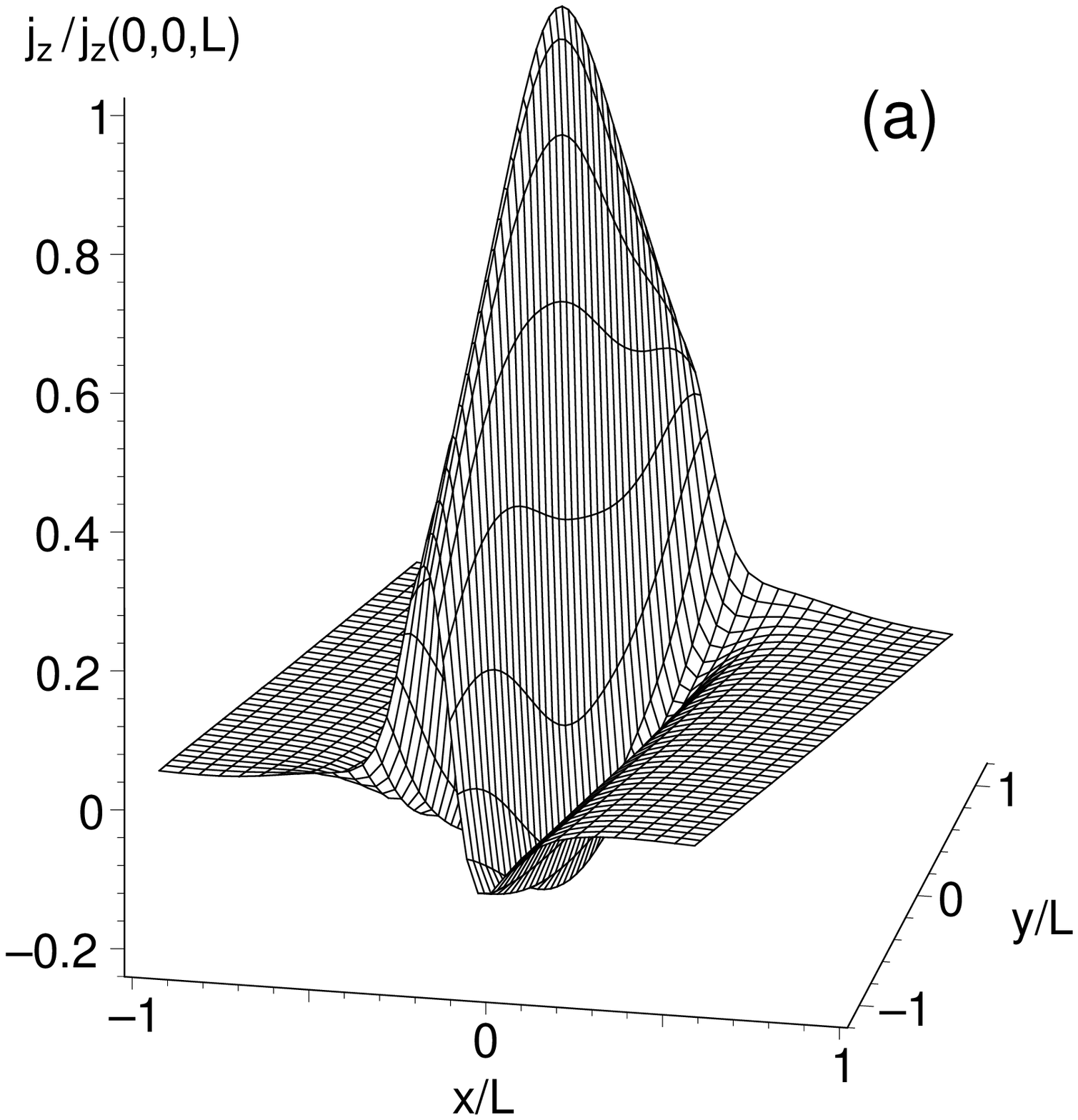}\\
\plotone{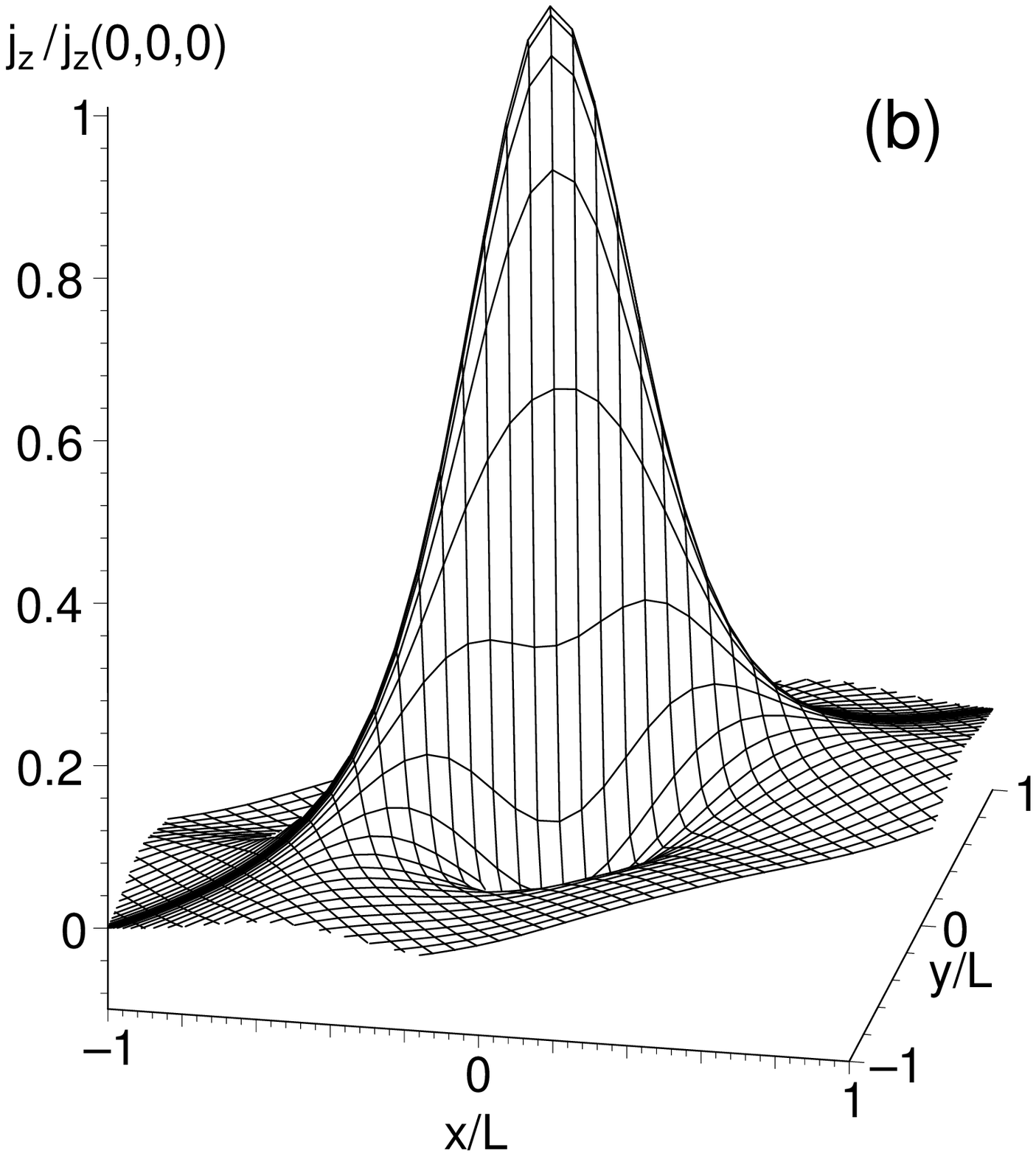}\\
\plotone{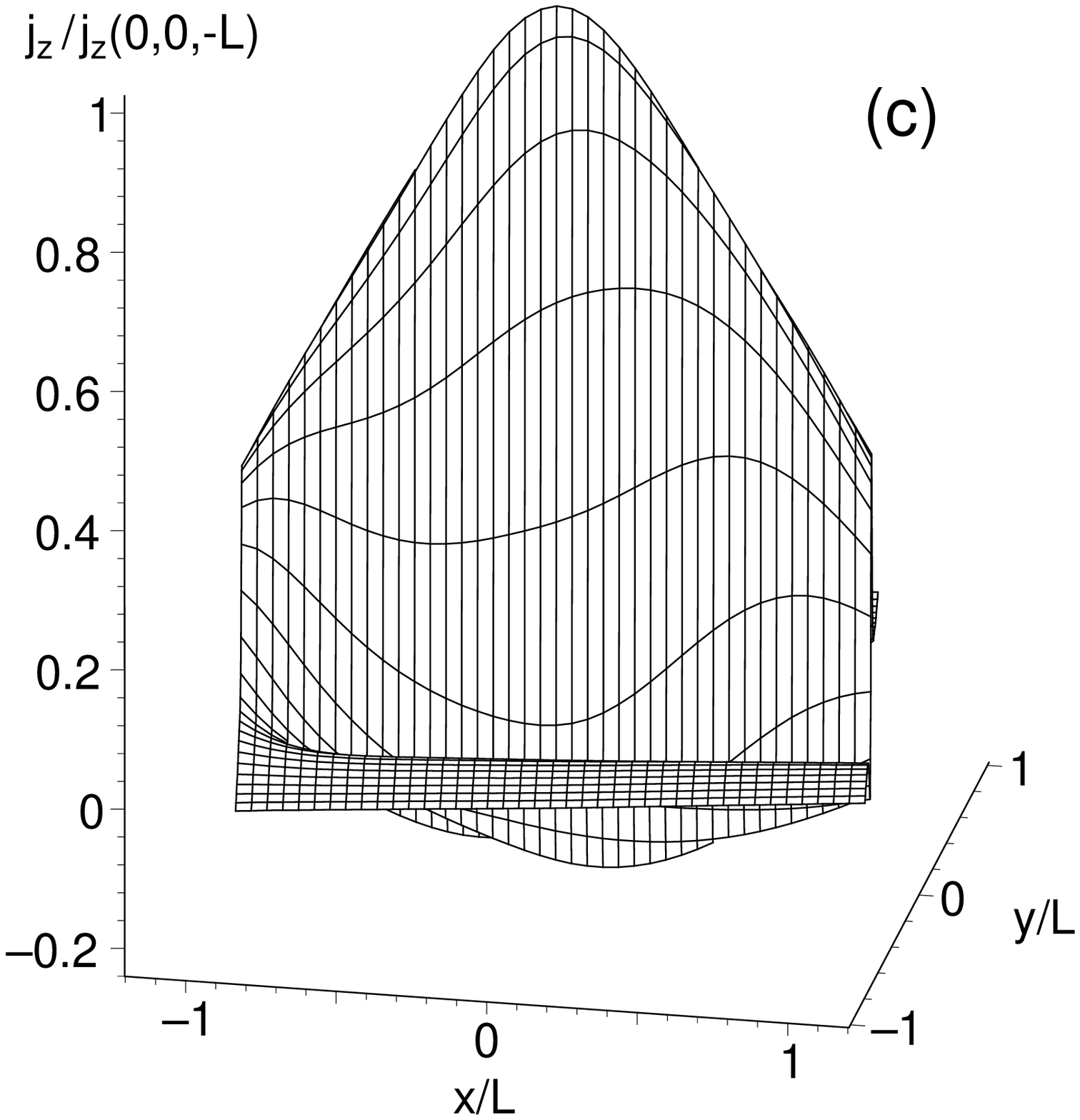}
\caption{The distribution of the current density $j_{z}$ in the planes
$z=L$ (a), $z=0$ (b) and $z=-L$ (c) at $\bar{t}=2.5$ for the deformation
due to the ``twisting'' photospheric shearing motion ($d/L=0.4$ and
$l/L=0.5$). 
 \label{f:jzz0} }
\end{figure}

An important question we have not addressed yet is what
impact the
pinching has on the current density within the HFT.
We start by computing the
$z$-component of the current density ${\bf j}$ in the midplane $z=0$ and
the photospheric planes $z=\pm L$ in our so far the purely kinematic
model of the pinching process. As the velocity field is known the
calculation can be carried out using the induction equation and
Amp\'ere's law. 
The results are shown in Figure~\ref{f:jzz0} for the time
$\bar{t}=2.5$.
A comparison of Figures~\ref{f:QL} and~\ref{f:jzz0}b shows that the
distributions of $j_{z}$ and $Q_{\rm L}$ in the plane $z=0$ have
nearly the same shapes.
The $j_{z}$ distribution becomes noticeably different from 
the $Q_{\rm L}$ distribution only in the
inflow region near the current layer, where spikes of reverse
current appear.
Such a similarity between the two distributions indicates that 
in this kinematic picture
the current layer forms as a direct consequence of the strong squashing of
Lagrangian elements.
The presence of the reverse currents in the inflow regions is very
interesting as similar features 
are also found in exact \citep{sonpr75,crhen95, prea00}
and in numerical \citep{bsk86} investigations of 2D reconnection.
It is a nice surprise to see that these features are reproduced in our
3D kinematic model. 
The reverse currents do not only exist in the midplane ($z=0$), but 
also in the planes $z=\pm L$ where they have
a slightly asymmetric form (Figure~\ref{f:jzz0}a,~c).
The current layer in these planes is mainly produced by 
the shearing flows along the
$y$-axis ($z=L$) or the $x$-axis ($z=-L$), which are induced by given
velocity field. 
In consequence, the current layer is wider, 
and longer than in the midplane, and nearly parallel to one of 
the coordinate axes.
Owing to symmetry of the initial configuration and 
of the applied boundary motions
(see Figures~\ref{f:emgs}, \ref{f:hft_str} and \ref{f:dfms}) 
the spatial distribution of $j_{z}$
in the planes $z=L$ and $z=-L$ is very similar except that
for $z=-L$ the layer is aligned with the $x$-axis whereas
for $z=L$ it is aligned with the $y$-axis.
Thus the orientation of the current layer is twisted by 
approximately $90^{\circ}$ when
moving from the lower to the upper boundary plane. 

Further interesting properties of the
pinching process can be learned from studying the time evolution of 
the magnetic
field and the current density distribution in the plane $z=0$ along the
line $y=-x$. 
This is most conveniently carried out in a new ($(x^{\prime}, 
y^{\prime})$) coordinate system in which one of the axis ($x^{\prime}$)
is aligned with
the current layer.
This coordinate system is rotated by $45^{\circ}$ with respect 
to the original one.
 The detailed derivation of the magnetic field component $B_{x^{\prime}}$ 
is described
in Appendix~\ref{s:conv}, while here we only present the resulting
expression: 
 \begin{eqnarray}
  B_{x^{\prime}} &=& B_{x^{\prime}1} + B_{x^{\prime}2}, 
     \label{Bxp} \\
B_{x^{\prime}1} &=& -\sqrt{2}\, h l_{\rm sh} e^{\bar{t}} \cosh \left(
    {\bar{y}^{\prime} \over \sqrt{2} } \right) {\arcsinh \left[
      e^{\bar{t}} \sinh \left( {\bar{y}^{\prime} \over \sqrt{2}} \right)
    \right] \over  \left[ 1 +  \left( e^{\bar{t}} \sinh \left(
          {\bar{y}^{\prime} \over \sqrt{2}
          }\right) \right)^{2} \right]^{1/2}}, 
    \label{Bxp1} \\
B_{x^{\prime}2} &=&-{B_{\|} \over \sqrt{2} }{l_{\rm sh} \over L}\,
      \cotanh  \left( { 
       \bar{y}^{\prime} \over \sqrt{2}} \right) \ln  \left[ {1 +  \left(
         e^{\bar{t}} \sinh \left( {\bar{y}^{\prime} \over \sqrt{2} }
         \right) \right)^{2} \over 1 + \sinh^{2}  \left(
         {\bar{y}^{\prime} \over \sqrt{2}} \right)} \right].
     \label{Bxp2}
 \end{eqnarray}
The $B_{x^{\prime}1}$ contribution is caused by 
the  stretching of the transverse magnetic field
along the $x^{\prime}$-axis due to the hyperbolic structure of the flow
in the middle of the HFT.
The $B_{x^{\prime}2}$ contribution appears because of 
the  $B_{z}$-component is sheared
by a non-vanishing gradient of the flow in the $z$-direction.
Notice that this second effect, contrary to the first one, is
purely three-dimensional.

\begin{figure}
\epsscale{0.4}
\plotone{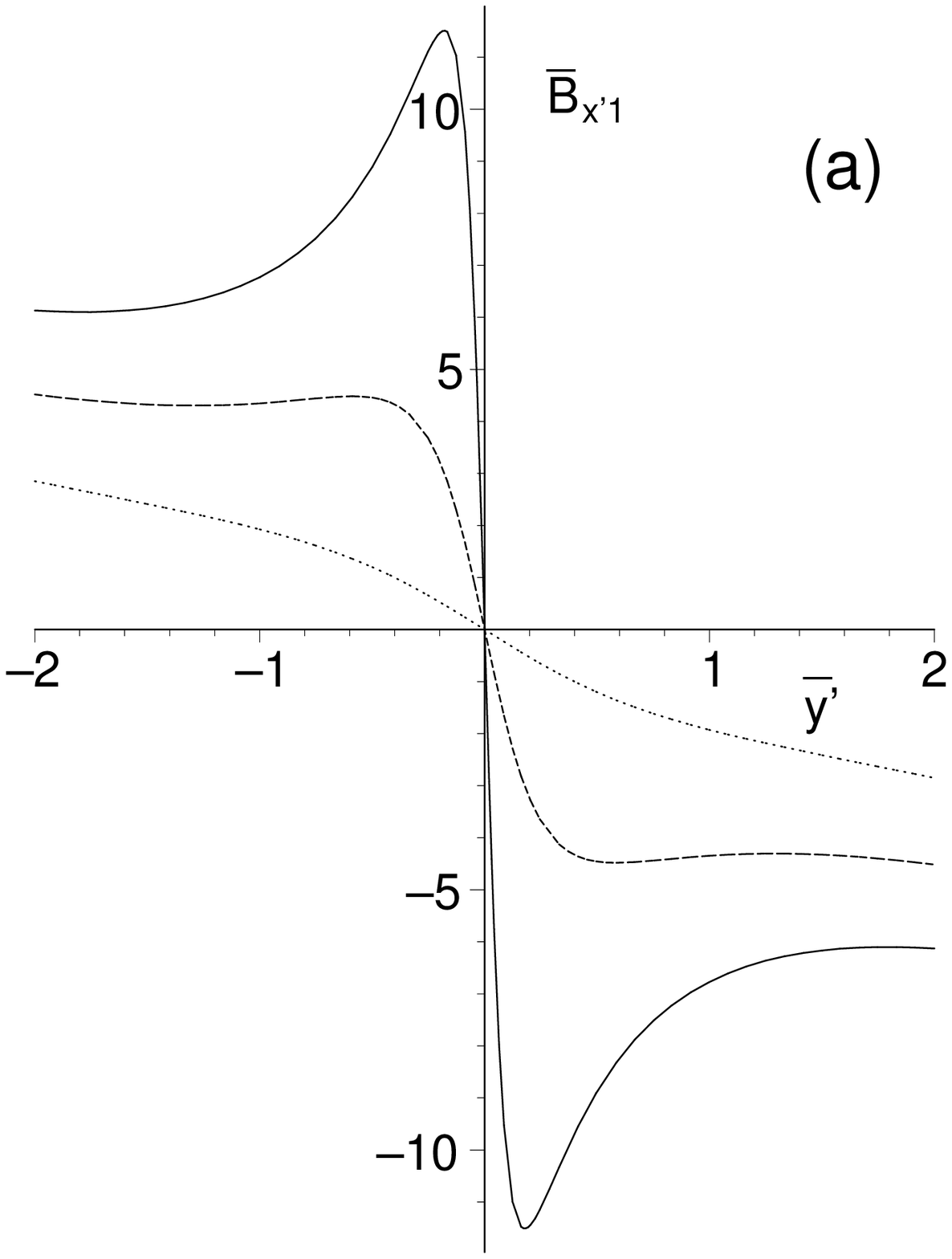}\plotone{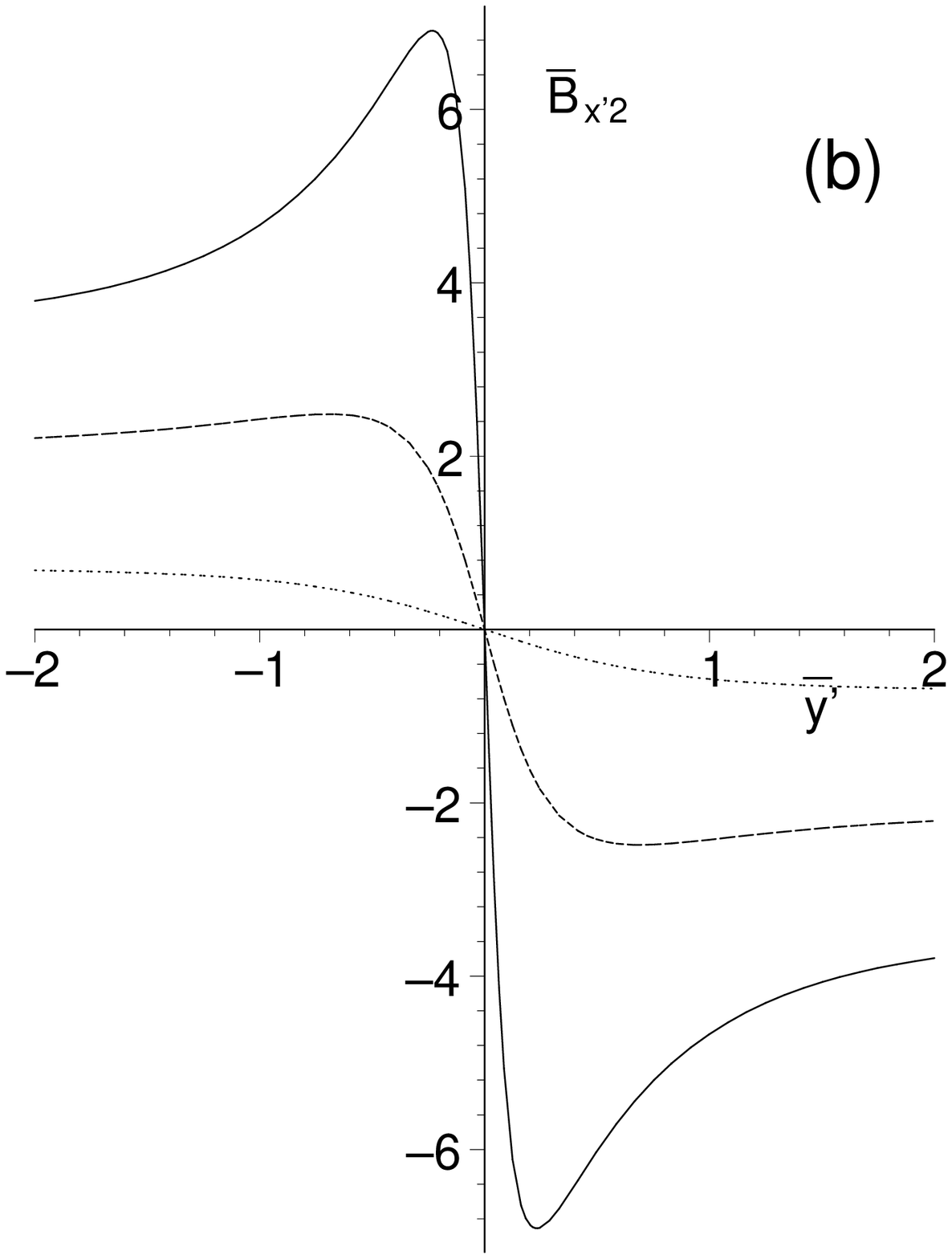}\\
\plotone{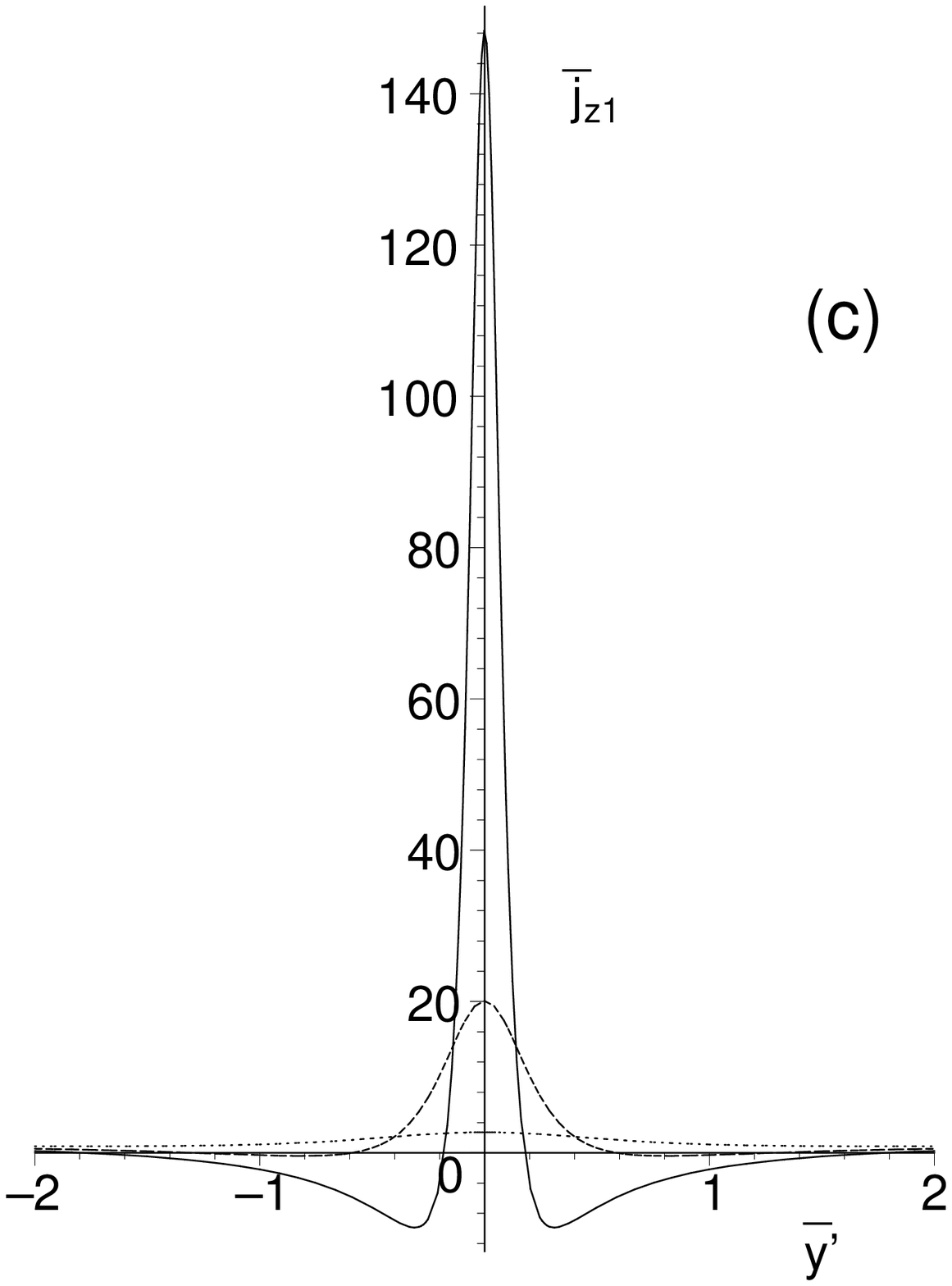}\plotone{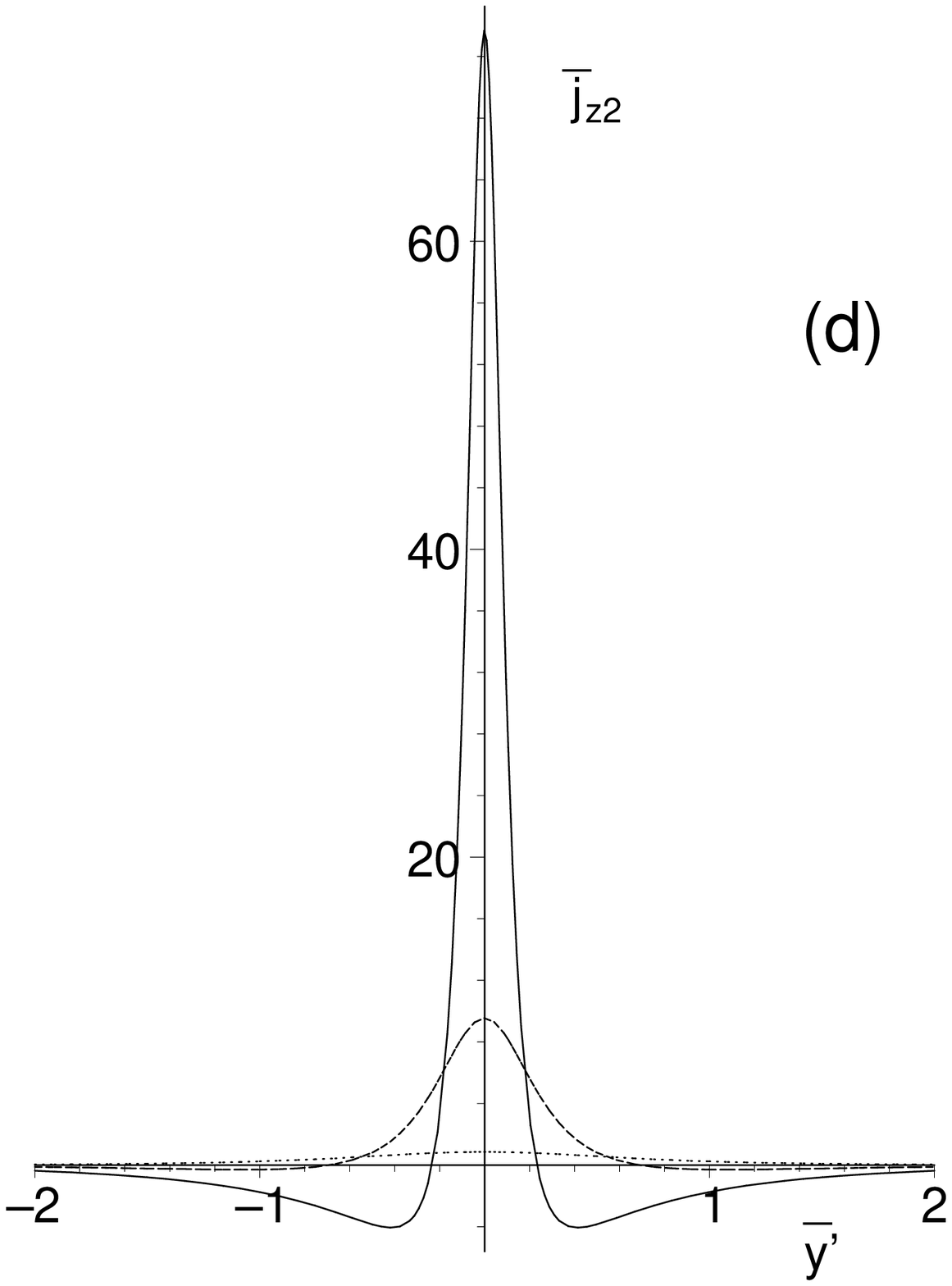}
\caption{The magnetic fields $\bar{B}_{x^{\prime}1}$ and
$\bar{B}_{x^{\prime}2}$ (a, b) normalized
to $h l_{{\rm sh}}$ and $B_{\|} l_{{\rm sh}}/L$, respectively,
and the corresponding current densities $\bar{j}_{z1}$ and $\bar{j}_{z2}$
(c, d) normalized to $h/\mu_{0}$ and $B_{\|}/(\mu_{0} L)$, respectively, at
$x^{\prime}=z=0$.  Dotted, dashed and solid lines correspond to
$\bar{t}=0.5$, $1.5$ and $2.5$, respectively.
 \label{f:prfls} }
\end{figure}

Despite this difference
the spatial and time variation of $B_{x^{\prime}1}$ and $B_{x^{\prime}2}$ 
are quite similar.
 In both cases a steepening of the profiles occurs at
$\bar{y}^{\prime}=0$ and the field is piled up at $\bar{t}\gtrsim 1.5$
close to the steepening fronts (cf. Figures~\ref{f:prfls}a, b).
 This means that the corresponding profiles of the current
densities $j_{z1} \approx -{\partial B_{x^{\prime}1} \over \partial
  y^{\prime}}$ and $j_{z2} \approx -{\partial B_{x^{\prime}2} \over
  \partial y^{\prime}}$ have growing and narrowing positive peaks at
$\bar{y}^{\prime}=0$ which for
$\bar{t}\gtrsim 1.5$ are enclosed by extended regions of reverse current
(Figures~\ref{f:prfls}c,~d).
 With the help of (\ref{Bxp})--(\ref{Bxp2}) one can find the maximum current
density $j_{z}^{*}$ at $\bar{t}\gtrsim 1$ given by
 \begin{eqnarray}
  j_{z}^{*} \equiv
  j_{z}\big|_{{\bf r}=0} \simeq e^{2\bar{t}} {h \over \mu_{0} } 
  \left(1+{B_{\|}\over 2 h L}\right),
  \label{jzX}
 \end{eqnarray}
which shows that for sufficiently large displacements of the spots
the kinematic HFT pinching causes an exponential
growth of the current density in the middle of the HFT\@.
 An exact form of $j_{z}^{*}$ (see below) contains
the factor $2\sinh(2\bar{t})$
instead of $e^{2\bar{t}}$, which
gives the correct linear growth of $j_{z}^{*}$ at $2\bar{t}\lesssim 1$ as
expected.

Appendix~\ref{s:conv} presents the technique which allows the derivation of
the $j_{z}$-distribution along the complete quasi-separator. 
Using this technique we obtain
 \begin{eqnarray}
  j_{z\,{\rm qs}} & = & j_{z\,{\rm qs}1} + j_{z\,{\rm qs}2} 
  \label{jzqs}, \\
    j_{z\,{\rm qs}1} & = & {\sinh(2\bar{t}\sqrt{1-\bar{z}^{2}}) \over
       \sqrt{1-\bar{z}^{2}}} {1 \over \mu_{0} }
      \left.\left({\partial B_{0x} \over \partial  x}
         -{\partial B_{0y} \over \partial  y}
       \right)\right|_{\rm qs} , 
    \label{jzqs1} \\
   j_{z\,{\rm qs}2} & = & \left[ \sinh(2\bar{t}\sqrt{1-\bar{z}^{2}}) -
  2\bar{z}^{2}\bar{t}\sqrt{1- \bar{z}^{2}} \right]
     {B_{0z}\big|_{{\rm qs}}/(\mu_{0}L) \over (1-\bar{z}^{2})^{3/2}}.
     \label{jzqs2}
 \end{eqnarray}
Here the index ${\rm qs}$ means that the corresponding magnitudes are
evaluated at ${\bf r} = (0,0,z)$ resulting in
 \begin{eqnarray}
  \left.\left({\partial B_{0x} \over \partial  x}
         -{\partial B_{0y} \over \partial  y}
       \right)\right|_{\rm qs} & = &
    6 \bar{d}^{2} \bar{l}^{2} {B_{{\rm s}}\over L} \left\{  \left[
        \bar{l}^{2} +   \left( 1 + \bar{d} +
     \bar{z} \right)^{2} \right]^{-5/2} + \right. 
  \nonumber \\
 &&\qquad\qquad\quad \left.  \left[ \bar{l}^{2} +  \left( 1 + \bar{d}
       - \bar{z} \right)^{2}  \right]^{-5/2}
 \right\}, 
 \label{dBqs} \\
 B_{0z}\big|_{{\rm qs}} & = &
  2 B_{{\rm s}}\bar{d}^{2} \left\{ {1 + \bar{d} +
     \bar{z}  \over
 \left[ \bar{l}^{2} + (1+\bar{d}+\bar{z})^{2} \right]^{3/2} } +
 \right. 
  \nonumber \\
   &&\qquad\qquad \left. {1 + \bar{d} - \bar{z} \over  \left[
         \bar{l}^{2} + (1 + \bar{d} -  \bar{z})^{2} 
 \right]^{3/2}} \right\}.
 \label{Bzqs}
 \end{eqnarray}
In the derivation of (\ref{jzqs})--(\ref{jzqs2}) we also used that
$\left.{\partial B_{0y} 
    \over \partial x}\right|_{{\rm qs}} =\left.{\partial B_{0x} \over
    \partial y}\right|_{{\rm qs}} \equiv 0$ for
our configuration~(\ref{B0}).  

Putting $\bar{z} \rightarrow \mp 1$ in (\ref{jzqs})--(\ref{jzqs2}),
one obtains $j_{z}$ at the footpoints of the quasi-separator
 \begin{eqnarray}
  j_{z\, {\rm qsf}} =
   {2\bar{t}\over \mu_{0}} \left.\left( {\partial B_{0x} \over \partial  x}
   -{\partial B_{0y} \over \partial  y}
    +{B_{0z}\over L} \right)\right|_{{\rm qsf}} 
 +{4 \bar{t}^{3} \over 3\mu_{0} L} B_{0z}\big|_{{\rm qsf}},
    \label{jzqsf}
 \end{eqnarray}
where the index ${\rm qsf}$ denotes the evaluation of the
corresponding magnitudes at ${\bf r}=(0,0,\mp L)$.
 Thus, the current density at the feet of the HFT grows 
only with the third power of time, 
which is much slower than the exponential growth in
the middle of the HFT\@.
 This difference is due to the fact that motions at the HFT
feet (\ref{vdwn})--(\ref{vup}) just shear the magnetic field. Such
a pure shearing motion is less effective in generating
current than the ``hyperbolic superposition'' in the middle of the HFT.

\begin{figure}
\epsscale{0.4}
\plotone{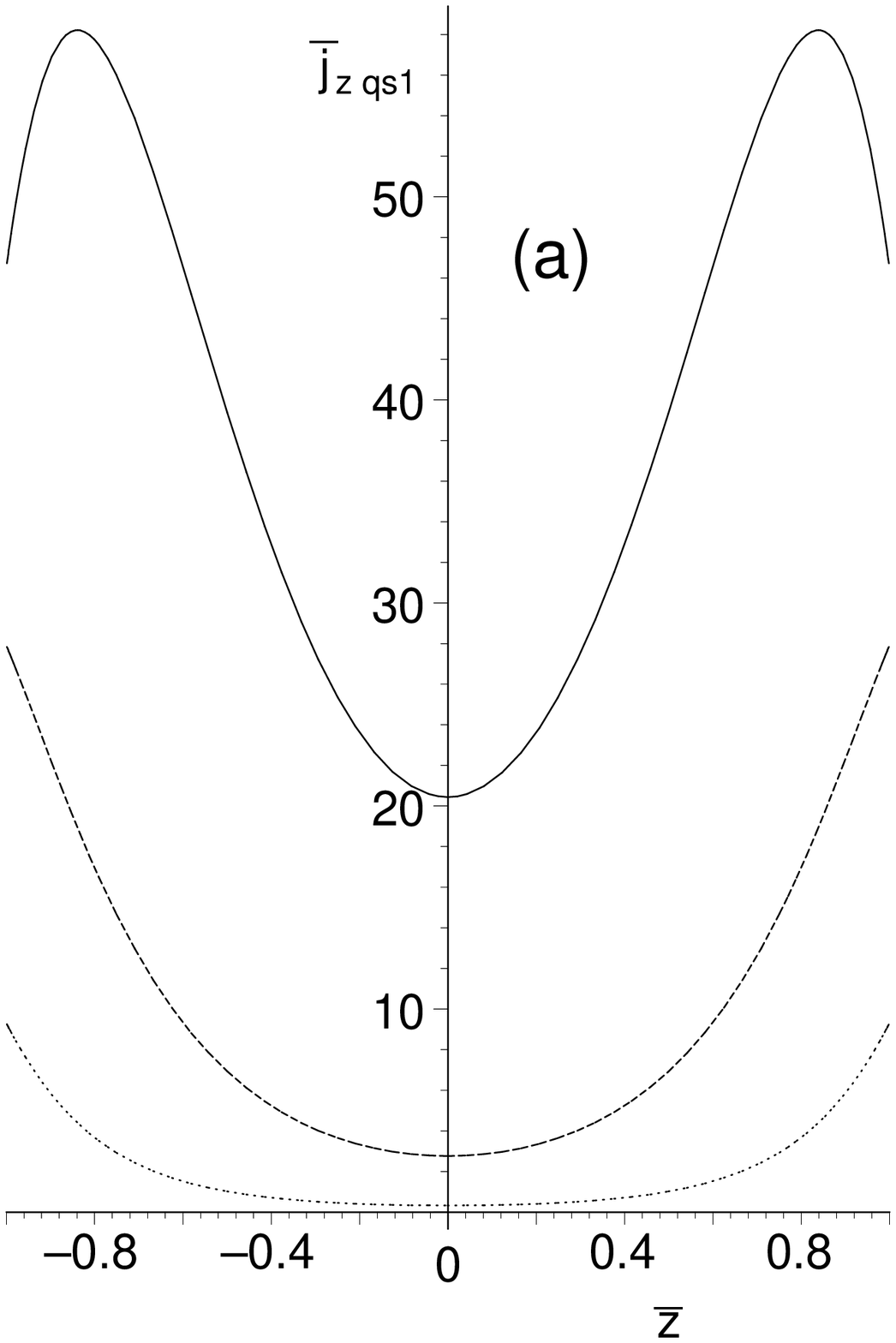}\\
\plotone{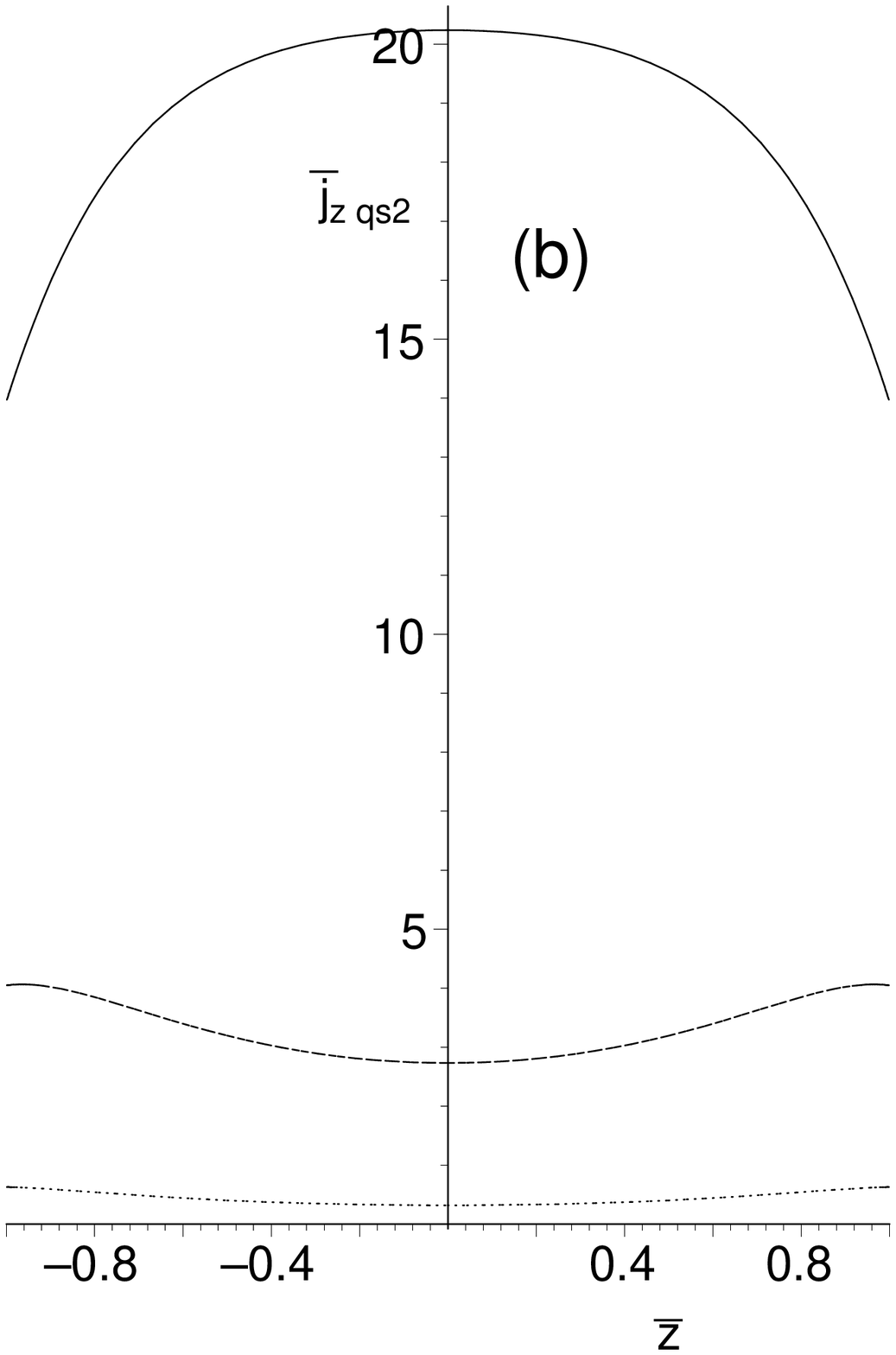}
\caption{The distributions of current densities $\bar{j}_{z\,{\rm
      qs}1}$ and $\bar{j}_{z\, {\rm qs}2}$ normalized to
$B_{{\rm s}}/(\mu_{0}L)$ along the quasi-separator ($\bar{z}=z/L$)
for the same parameter values as used in Figure~\ref{f:emgs}. Dotted, dashed
and solid lines correspond to $\bar{t}=0.5$, $1.5$ and $2.5$, respectively.
 \label{f:jzqs} }
\end{figure}

However, the local gradients of magnetic field are much larger at the feet
than in the middle, therefore the maxima of $j_{z\,{\rm qs}1}$
remain located at the feet even for the times during which the
sunspots have moved over distances comparable with $L$
(Figure~\ref{f:jzqs}a).  
 By contrast, the second contribution to the current density 
($j_{z\,{\rm qs}2}$) has a maximum in the middle of the HFT 
at comparable times (Figure~\ref{f:jzqs}b)
because of the above mentioned exponential growth and a relatively
small gradient of $B_{0z}$ along the quasi-separator.

\section{FORCE-FREE PINCHING OF THE HFT}
\label{s:ffp}

In the kinematic model of the HFT pinching process we considered so far,
 the force balance
is taken into account only qualitatively by constructing
deformations of the HFT which are in agreement with its elastic properties.
 However, the layer-like structure obtained in the kinematic model for the
middle part of the HFT enables us to improve the model by determining the
distributions of the field and current density across such a
layer for a force-free evolution of the configuration.

\begin{figure}
\epsscale{0.4}
\plotone{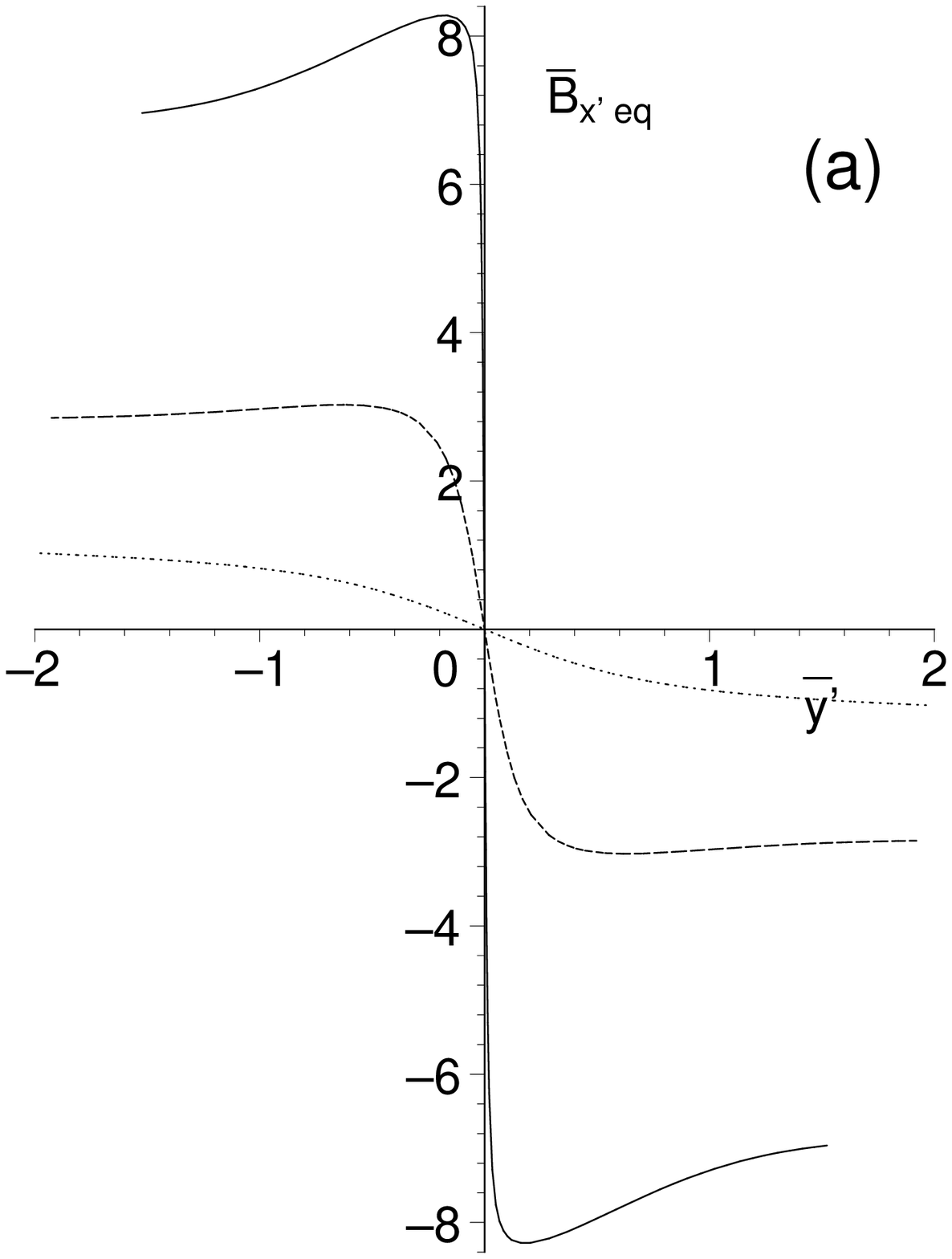}\\
\ \\
\plotone{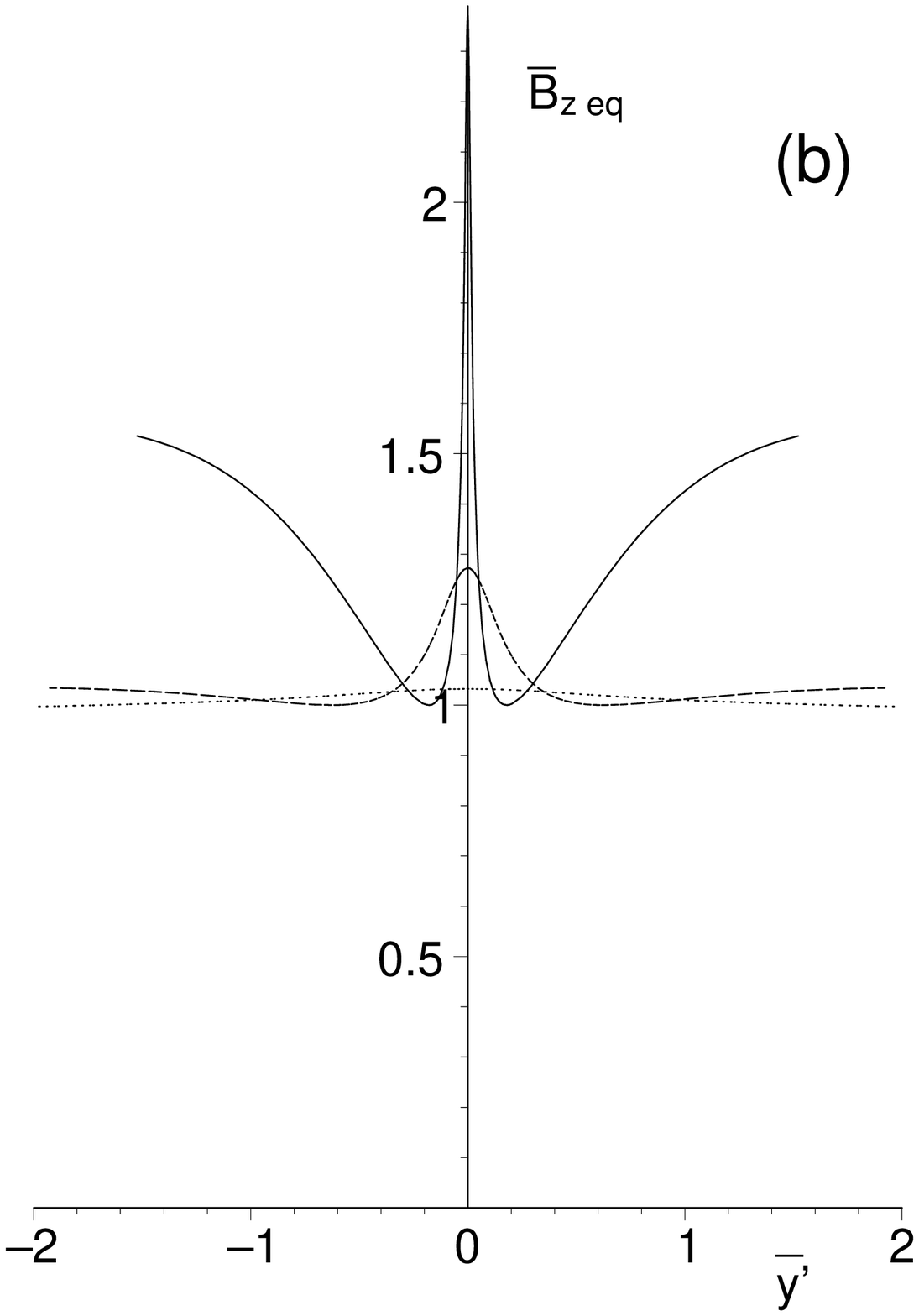}
\caption{The profiles of the equilibrium field components $B_{x^{\prime}\,
    {\rm eq}}$ and $B_{z\,{\rm eq}}$ normalized to
$B_{\|} l_{{\rm sh}}/L$ and $B_{\|}$, respectively,
for the same parameter values as used in Figure~\ref{f:emgs}. Dotted, dashed
and solid lines correspond to $\bar{t}=0.5$, $1.5$ and $2.5$, respectively.
 \label{f:Beq} }
\end{figure}

Indeed, at each time such a force-free evolution can be represented as a
deformation of the initial potential field obeying the frozen-in
condition.
One can always decompose such a deformation into
the kinematic ``pinching'' deformation considered so far
 and an additional ``relaxing'' deformation in which
the configuration settles into a force-free state. 
 As shown above, after the pinching deformation unbalanced
magnetic stresses accumulate in the middle of the HFT, mainly in a thin
layer parallel to the plane $y^{\prime} = 0$.
 This is due to an increase of the $B_{x^{\prime}}$-component
and the magnetic pressure it causes near this plane, 
so that the corresponding unbalanced
force is directed towards the plane acting as if to compress the layer.
We conclude that the relaxing deformation will compress
the middle part of the HFT\@.
 The layer-like distribution of the unbalanced stresses implies that this
compression will be essentially one-dimensional.
Since $B_{z}$, unlike $B_{x^{\prime}}$,
does not vanish inside the layer, the corresponding
compression is indeed able to provide the required balance of magnetic
forces inside the layer.
With the expressions (\ref{Bxp})--(\ref{Bxp2}) for $B_{x^{\prime}}$ and 
taking 
$B_{\|}$ for $B_{z}$ as initial values, one can use the frozen-in 
condition (see (\ref{BL})) to obtain the components of the
equilibrium field as
 \begin{eqnarray}
  B_{x^{\prime}\, {\rm eq}} & \simeq & 
  B_{x^{\prime}}\big|_{\bar{y}^{\prime}=\bar{y}^{\prime}_{0}}
    \left( 
   {\partial \bar{y}^{\prime} \over \partial \bar{y}^{\prime}_{0}}
   \right)^{-1},
   \label{Bxpe} \\
  B_{z\,{\rm eq}} & \simeq & B_{\|} \left( 
  {\partial \bar{y}^{\prime} \over \partial \bar{y}^{\prime}_{0}}
  \right)^{-1}. 
  \label{Bze}
 \end{eqnarray}
These magnetic field components have to satisfy the force-free condition
 \begin{eqnarray}
  B_{x^{\prime}\, {\rm eq}}^{2} + B_{z\,{\rm eq}}^{2} \simeq \mbox{const},
  \label{ffc}
 \end{eqnarray}
which together with (\ref{Bxpe}) and (\ref{Bze}) yields the equation for
determining the relaxing deformation
$\bar{y}^{\prime}(\bar{y}^{\prime}_{0})$ or, in other words, the
relationship between Eulerian ($\bar{y}^{\prime}$) and Lagrangian coordinates
($\bar{y}^{\prime}_{0}$). 
 The constant in (\ref{ffc}) is naturally fixed by the condition that
$\partial \bar{y}^{\prime}/\partial \bar{y}^{\prime}_{0} = 1$ at the point
$\bar{y}^{\prime}_{0}=\bar{y}^{\prime}_{0{\rm m}}$, where the initial
current density and the unbalanced Lorentz force vanish and so the relaxing
deformation should have no effect on the neighborhood of this location. 
Therefore, at this point
$\partial B_{x^{\prime}}/\partial \bar{y}^{\prime}_{0} \approx 0$ and
$B_{x^{\prime}}^{2}$ reaches a local maximum $B_{x^{\prime}{\rm
    m}}^{2}$. 
 This makes it possible to derive from
(\ref{Bxpe})--(\ref{ffc}) the required equation
 \begin{eqnarray}
  {\partial \bar{y}^{\prime} \over \partial \bar{y}^{\prime}_{0}} =
 \left( 
   1 + B_{x^{\prime}}^{2}\big|_{\bar{y}^{\prime}=\bar{y}^{\prime}_{0}} /
     B_{\|}^{2}
        \over
  1 + B_{x^{\prime}\, {\rm m}}^{2} / B_{\|}^{2}
 \right)^{1/2},
   \label{dydy0}
 \end{eqnarray}
where $B_{x^{\prime}\, {\rm m}}^{2}$ depends on time
and the parameters of the model as shown below.
At a fixed time this is an ordinary differential equation for
$\bar{y}^{\prime}(\bar{y}^{\prime}_{0})$, which can be immediately 
integrated to give
 \begin{eqnarray}
   \bar{y}^{\prime} = \int_{0}^{\bar{y}^{\prime}_{0}}  
 \left( 
   1 + B_{x^{\prime}}^{2}\big|_{\bar{y}^{\prime}=s} /
     B_{\|}^{2}
        \over
  1 + B_{x^{\prime}\, {\rm m}}^{2} / B_{\|}^{2}
 \right)^{1/2} {\rm d}s.
 \label{ype}
 \end{eqnarray}
 By differentiating (\ref{Bxpe}) and (\ref{dydy0}) (to obtain an
expression for $\partial^{2} \bar{y}^{\prime} / {\partial
    \bar{y}_{0}^{\prime}}^{2}$) one can also derive
the equilibrium current density,
 \begin{eqnarray}
  j_{z\,{\rm eq}}  \simeq  -{1 \over \mu_{0} l_{{\rm sh}}}
  \left. {\partial B_{x^{\prime}} \over \partial  \bar{y}^{\prime}}
  \right|_{\bar{y}^{\prime}=\bar{y}^{\prime}_{0}}
{1 + B_{x^{\prime}\, {\rm m}}^{2} / B_{\|}^{2} \over \left(
   1 + B_{x^{\prime}}^{2}\big|_{y^{\prime}=\bar{y}^{\prime}_{0}} /
     B_{\|}^{2}\right)^{2}}.
  \label{jze}
 \end{eqnarray}
 Thus, Eqs. (\ref{Bxpe}), (\ref{Bze}) and
(\ref{dydy0})--(\ref{jze}) determine 
the distributions of the equilibrium magnetic field and
current density across the current layer in the middle of
the HFT in parametric form with
$\bar{y}^{\prime}_{0}$ as a parameter.
 Examples of the $B_{x^{\prime}\, {\rm eq}}$- and 
$B_{z\,{\rm eq}}$-profiles at different times are shown in
Figure~\ref{f:Beq}.
The plots show that the steepening of $B_{x^{\prime}\, {\rm eq}}$ in time
is accompanied by a corresponding compression of $B_{z\,{\rm eq}}$
inside the current layer.
 
For the especially interesting large time regime ($\bar{t}\gtrsim 1$)
one can show that two extrema of $B_{x^{\prime}}$ are located at
 \begin{eqnarray}
  \bar{y}^{\prime}_{0{\rm m}} \simeq  
  \pm{\delta \over 2\, l_{{\rm sh}}}
   F\left({B_{\|} \over hL} \right). 
  \label{ym}
 \end{eqnarray}
In this expression $F$ is a relatively complicated 
function of its argument which we do not write down
explicitly here, but it turns out that
$F$ only increases from $1.51$ to $1.98$ when its argument
$B_{\|}/hL$ varies from 0 to infinity, so that to a good
approximation (\ref{ym}) is given by
 \begin{eqnarray}
  \bar{y}^{\prime}_{0{\rm m}} \simeq \delta/l_{{\rm sh}}.
  \label{yma}
 \end{eqnarray}
So an estimate of $B_{x^{\prime}\, {\rm m}}$ can be obtained by evaluating 
(\ref{Bxp}) at $\bar{y}^{\prime} =
\bar{y}^{\prime}_{0{\rm m}}$ under the assumption $\bar{t}\gtrsim
1$ with the result
 \begin{eqnarray}
 {B_{x^{\prime}\, {\rm m}}^{2} \over B_{\|}^{2}} \simeq
 e^{2\bar{t}}  \left( 0.91{hl_{{\rm sh}} \over B_{\|}}
   + 0.57{l_{{\rm sh}} \over L} \right)^{2}.
 \label{B2r}
 \end{eqnarray}  
 Substitution of this value and
$B_{x^{\prime}}\big|_{\bar{y}^{\prime} = 0} =0$ into
(\ref{jze}) enables us to express
the equilibrium current density at ${\bf r}=0$
in terms of the corresponding parameters of the initial configuration,
 \begin{eqnarray}
  j_{z\, {\rm eq}}^{*} \simeq
  e^{2\bar{t}} {h \over \mu_{0} } 
  \left(1+{B_{\|}\over 2 h L}\right)
  \left[ 1 + 
  e^{2\bar{t}} \left( 0.91{h l_{{\rm sh}} \over B_{\|}}
 + 0.57 {l_{{\rm sh}} \over L}  \right)^{2}
  \right].
   \label{jzeX}
 \end{eqnarray}
It follows that the ``relaxing'' compression
gives rise to an additional term in the
force-free current density in comparison with its kinematic analogue
(\ref{jzX}). This term grows exponentially
in time and increases with decreasing~$B_{\|}$.
 In particular, in the limit $B_{\|} \rightarrow 0$ we would formally obtain
that $|j_{z\, {\rm eq}}^{*}| \to \infty$, which is in good
agreement with the results for current accumulation at a null
point \citep{bulol84, prtit96, rcktit96}.

\section{IMPLICATIONS FOR SOLAR FLARES}
\label{s:impl}

\begin{figure}
\epsscale{0.6}
\plotone{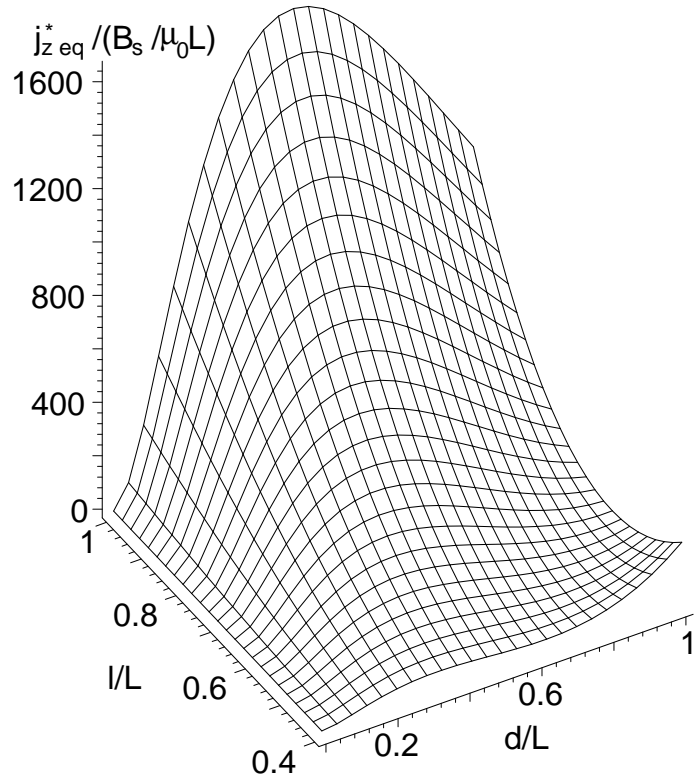}
\caption{The equilibrium current density $j_{z\ {\rm eq}}^{*}$ at
  ${\bf r}=0$ and $\bar{t}=2.5$ versus the model parameters $l/L$
  and $d/L$.
 \label{f:jzeq} }
\end{figure}

Figure~\ref{f:jzeq} shows the dependence of the current density (\ref{jzeX})
 on the geometrical
parameters $l/L$ and $d/L$ of our ``straightened'' HFT
for the time $\bar{t}=2.5$, where $h$,
$B_{\|}$ and $l_{{\rm sh}}$ are given by (\ref{h}), (\ref{Bpar}) and
(\ref{lsh}), respectively.
The largest values of $j_{z\ {\rm eq}}^{*}$ at this time are reached if
$l/L\rightarrow 1$ and $d/L\approx 0.5$ and they can exceed the
characteristic current density $B_{{\rm s}}/(\mu_{0}L)$ in the
configuration by more than a factor $10^{3}$. 
 Higher values of $j_{z\ {\rm eq}}^{*}$ can formally be reached at
larger $\bar{t}$ but they do not seem to be realistic, because
it is more likely that the current layer will be disrupted by
the onset of the tearing instability before that.

We evaluate the characteristics of the pinched HFT at the time
$\bar{t} = 2.5$ for the same
parameters ($l/L=0.5$ and $d/L=0.4$) as used in Figure~\ref{f:emgs}.
 Assume that $L=50\:{\rm Mm}$ and $B_{{\rm s}}=800\:{\rm G}$, then
for our ``straightened'' model HFT we obtain that the transverse and 
longitudinal fields  near the current
layer are $B_{x^{\prime}{\rm
    m}} \simeq 470\:{\rm G}$ and $B_{\|} \simeq 218\:{\rm G}$, respectively.
 Inside the layer  $B_{z}\big|_{{\bf r}=0} \simeq 518\:{\rm G}$
due to the compression of the longitudinal field by relaxation,
so that the corresponding compression factor is about~2.8.

To get a rough estimate of the free magnetic energy associated with
the current layer, one can use the 2D solution found by \citet{syr81}
as an approximation, 
because the current density in our model is mainly concentrated in a thin 
layer of width
$\Delta$ and length $2L$.
  Then in our notation the total current in the layer is
 \begin{eqnarray}
  I \simeq {\pi \over 2\mu_{0}} B_{x^{\prime}{\rm m}} \Delta,
  \label{I}
 \end{eqnarray}
while the magnetic flux associated with the current is given by
 \begin{eqnarray}
  {\cal L} I \simeq { 1 \over 4}
     \left( 1 + \ln{8 L \over \Delta} \right)
   B_{x^{\prime}{\rm m}}\Delta,
     \label{LI}
 \end{eqnarray}
where ${\cal L}$ is the inductance of the current sheet per unit
length.
The corresponding magnetic energy is 
 \begin{eqnarray}
  W \simeq {\cal L}I^{2} L \simeq {\pi \over 8\mu_{0}} 
     \left( 1 + \ln{8 L \over \Delta} \right)
  B^{2}_{x^{\prime}{\rm m}} \Delta^{2} L,
    \label{W}
 \end{eqnarray}
which for the parameter values given above yields $W \simeq
1.3\times 10^{26}\:{\rm J}$.
  This amount of energy is even more than needed to produce a large
flare and it is build up in our model over a reasonable time period of
$t=2.5\times 2l_{\rm
  sh}/V_{\rm s}\simeq 3.6\: {\rm h}$, if a sunspot
velocity $V_{\rm s} = 5\;{\rm km\, s^{-1}}$ is assumed.

For the same parameter values the maximum current density is 
given by $j^{*}_{z\,{\rm eq}}
 \simeq 0.33\:{\rm A}{\rm m}^{-2}$, which is not large enough for the
Spitzer resistivity to be important.
 Indeed, the local magnetic Reynolds number for our current layer
is
 \begin{eqnarray}
  Re_{m} = {V_{\rm s} B_{\rm s} \over \eta  j^{*}_{z\,{\rm eq}} } \simeq 44
  \gg 1,
  \label{Rem}
 \end{eqnarray} 
where the same sunspot velocity as above has been assumed and 
the Spitzer resistivity
$\eta$ is evaluated at the coronal temperature $T=2 \times 10^{6}
\:{\rm K}$.

The current density in such a layer is also not large enough for the
onset of the ion-acoustic instability, since the corresponding
critical current density $j_{\rm is} = e n c_{\rm is} \simeq 49\: {\rm
  A}\, {\rm m}^{-2}$ is much larger than $j^{*}_{z\,{\rm eq}}$.
In this estimate of $j_{\rm is}$ the evaluated compression factor is
taken into account, so that the plasma density inside the sheet
is taken to be $n=2.8 \times
10^{15}\: {\rm m}^{-3}$.
 The sound speed  $c_{\rm is}= \left(
 k_{\rm B} T/m_{\rm p} \right)^{1/2}$ is estimated for the 
coronal temperature given above.
 Other micro-instabilities
are characterized by similar or even larger critical values of the current
density \citep{som92, prfrb00} so that their development in our
current layer is also hardly possible. 

For the tearing instability the present
state of the theory, based on rather idealized models, does not allow
us to draw any reliable conclusions about its possible onset in the layer.
Intuitively, however, this seems to be quite a plausible possibility
and has some support from 2.5D numerical simulations of current layer
formation in quadrupole configurations \citep{hirea01}.
We believe that our 3D configuration therefore deserves careful
studies regarding this aspect in the future.
We will here assume that the current layer generated by the HFT
pinching process is unstable to the tearing mode
as a working hypothesis to see how this assumption
fits in with other known facts about flares.

In this sense we can say that the current layer in a pinched HFT will
be more unstable the larger the value of the current density
$j_{z\, {\rm eq}}^{*}$ is for given sunspot 
displacements.
 As mentioned above, $j_{z\, {\rm eq}}^{*}$ grows very fast with
decreasing $B_{\|}$, which means that HFTs with smaller
initial values of $B_{\|}$ will be more unstable.
 This fits in very well with the existence of two observationally
different classes of flares, the so-called less-impulsive and
more-impulsive flares \citep{sakea98}. 
 Such a distinct difference between the flares could be due to the 
strong dependence of $j_{z\, {\rm eq}}^{*}$ on $B_{\|}$ found above.
 A similar explanation has previously been proposed 
qualitatively by \citet{somkos97}.

We have demonstrated above that a ``twisting'' 
shearing motion at the HFT feet leads to the
pinching of its middle parts into a thin current layer 
with a free magnetic energy
sufficient for a large solar flare.
This kind of photospheric motion is in a good agreement with the
observational fact that large flares occur  frequently in the
configurations with an S-shaped photospheric polarity
inversion line \citep{mrtr66, morsev68}.
Such a relationship has been demonstrated first by \citet{gs88} in a
potential field model of a quadrupole configuration
using monopole sources.
These results have been confirmed and 
quantified generalizing the separator field line, on which
\citet{gs88} have based their analysis, to an HFT.

Another remarkable feature of the HFT pinching process considered
in this paper is 
the increase of the plasma density inside the current layer due to
the compression of the longitudinal field by relaxation under the
frozen-in condition.
 As can be seen from our estimates, this relaxation leads to a
compression of the coronal plasma inside the layer by up to one order
of magnitude, which may be
sufficient for the onset of thermal instability and, hence, for an
additional condensation of plasma in the current layer. 
Such a side effect of the pinching process has two important implications.

First, the accumulation of plasma inside the current layer
in the pre-flare state increases the number of electrons available
for acceleration in the subsequent impulsive phase of the flare when the
layer becomes unstable.
 This can help to solve the electron number problem of particle
acceleration in flares \citep{millea97}.
 The observational evidence  recently reported by \citet{klea02} provides
support for this point of view.

 Secondly, the compression and condensation of plasma in the pinching
HFT is a good candidate for a 3D mechanism of prominence formation
in the solar corona under suitable physical conditions.
 At present such a mechanism has only been investigated in the
framework of a 2.5D numerical model \citep{hirea01}.
Its generalization to the 3D geometry of the HFT might help to develop a
unified theory of solar flares and prominence formation, 
two phenomena which so far seemed to be  completely unrelated. 

\section{CONCLUSIONS}
\label{s:cs}

We have found that the fundamental condition for the formation of
layers of particularly strong current density (pinching) inside
hyperbolic flux tubes (HFTs) are photospheric shearing motions which
twist the HFT around its axis.
 Photospheric motions which merely turn the HFT have a much weaker effect.
 We plan to study in more detail in the following papers of this
series how combinations of these two extreme types of photospheric
motions influence the HFT pinching.
 Also the obtained results must be valid for configurations with
separator field lines and their associated separatrix surfaces, since
such structures are a limiting case of HFTs.

Due to  the special elastic properties of the HFT the photospheric
shearing motions easily propagate into the corona and meet each other
in the middle of the HFT\@. 
 If the shearing motions are applied to the HFT feet in a way as if
trying to cause a torsion of the HFT, their superposition in the
middle of the HFT forms a  hyperbolic flow pattern which is inclined
by $45^{\circ}$ with respect
to the hyperbolic structure of the transverse magnetic field inside
the HFT\@.
 Such an arrangement of the transverse velocity and magnetic fields
provides an exponentially growing current density 
ordered in a layer-like structure along the HFT\@.
 The width of the current layer is of the order of the characteristic
size of the shearing motions, while its thickness decreases
exponentially fast in time.

There are basically two physical effects which are responsible for the
process of HFT pinching.
 First, the hyperbolic component of the flow is 
incompressible and so does not change the longitudinal magnetic
field in the HFT\@. 
 However, it causes a large squashing of plasma elements
along the forming layer and this squashing increases the transverse
magnetic field near the layer enormously in comparison with the
initial state. 
Therefore, the strength of the resulting transverse field can easily
approach that of the longitudinal field and even exceed it, depending
on the initial conditions and the duration of the flow.
 The increase of the transverse field causes a second effect -- the
compression of the  longitudinal field and plasma across the layer,
which leads to an additional increase of the current density inside
the layer.

The exponential time dependence of the basic physical quantities inside
the current layer  can also be translated into
a similar dependence on sunspot displacements.
 For a quadrupole configuration with
an HFT  we can estimate that the displacements of the spots over
distances comparable with the distances between them are sufficient
to build up magnetic energy inside the pinched HFT as is
required for a large solar
flare.
 The large current density and gradient of magnetic shear inside the
layer provide favorable conditions for the onset of the tearing
instability in the layer and its transition to a turbulent state with
a large rate of reconnection and magnetic energy release.  

\acknowledgments

The authors are grateful to the anonymous referee for a number of
useful comments.
VT thankfully acknowledges financial support from the
Volkswagen-Foundation. KG and TN thank the UK's 
Particle Physics and Astronomy Research
Council for support through Advanced Fellowships.
This work has been supported in part by
the European Community's Human Potential
Programme under contract HPRN-CT-2000-00153, PLATON.

\appendix

\section{JACOBIAN MATRIX OF THE FIELD LINE MAPPING}

For an analytical investigation of the field line mapping in a given
magnetic configuration it is useful to have a method for calculating
the Jacobian matrix of this mapping.
Below such a method is developed and applied to calculate the
squashing degree at the quasi-separator of the ``straightened'' HFT
(see \S~\ref{s:S2}).

Let {\bf B}({\bf r}) define a magnetic field and
 \begin{eqnarray*}
  {\bf R}(\tau, {\bf r}_{+}) \equiv (X(\tau,{\bf
r}_{+}), Y(\tau, {\bf r}_{+}), Z(\tau, {\bf r}_{+}))
 \end{eqnarray*}
represent its field lines starting at points ${\bf
r}_{+}=(x_{+},y_{+})$ of the positive photospheric polarity located in
the plane $z=z_{+}$.
 The field lines are parameterized by a parameter $\tau$ such that the
corresponding field line equation is 
 \begin{eqnarray}
  {{\rm d}{\bf R} \over {\rm d}\tau} = {\bf B}({\bf R}).
        \label{fleq}
 \end{eqnarray}
The differentiation of (\ref{fleq}) with respect to ${\bf r}_{+}$ yields
 \begin{eqnarray}
   {{\rm d}{\cal D} \over {\rm d}\tau} = {\cal G}{\cal D},
        \label{Deq}
 \end{eqnarray}
where ${\cal G}\equiv \nabla_{\bf R} {\bf B}$ is the matrix
of derivatives of the magnetic field with respect to  ${\bf R} = {\bf
  R}(\tau, {\bf r}_{+})$, and
 \begin{eqnarray}
 {\cal D} \equiv {\cal D}(\tau, {\bf r}_{+})
=\nabla_{{\bf r}_{+}} {\bf R}(\tau, {\bf r}_{+}) \equiv
\left( \begin{array}{cc}
 {\partial X \over \partial x_{+}}
    & {\partial X \over \partial y_{+}}\\
 {\partial Y \over \partial x_{+}}
    & {\partial Y \over \partial y_{+}}\\
{\partial Z \over \partial x_{+}}
    & {\partial Z \over \partial y_{+}}
           \end{array}
    \right)
        \label{D}
 \end{eqnarray}
is the Jacobian matrix.
With its help a variation $\delta{\bf R}$ of
the field line point ${\bf R} = {\bf R}(\tau, {\bf r}_{+})$ due to
an infinitesimal change $\delta{\bf r}_{+}=(\delta x_{+}, \delta
y_{+})$ of the corresponding footpoint
${\bf R}_{+} = {\bf R}(0,{\bf r}_{+}) \equiv (x_{+},y_{+},z_{+})$ is
determined by
 \begin{eqnarray} 
   \delta{\bf R}={\cal D}\, \delta{\bf r}_{+}.
        \label{dR}
\end{eqnarray}

  The system of nine ordinary differential
Eqs. (\ref{fleq})--(\ref{Deq}) together with the initial 
conditions
 \begin{eqnarray}
  {\bf R}(0,{\bf r}_{+}) & =  &(x_{+},y_{+},z_{+}),
        \label{Rin} \\
  {\cal D}(0,{\bf r}_{+}) & = &
 \left( \begin{array}{cc}
    1 & 0 \\
    0 & 1 \\
    0 & 0
        \end{array}
  \right)
        \label{Din}
 \end{eqnarray}
determines the matrix ${\cal D}$ along the field line starting at a
point ${\bf r}_{+}$ on the photosphere.
  Thus the matrix ${\cal D}$ can be obtained as the result of
solving the Cauchy problem defined by Eqs. (\ref{fleq})--(\ref{Deq})
and (\ref{Rin})--(\ref{Din}).

  It should be emphasized that the vector $\delta{\bf r}_{+}$ in Eq. 
(\ref{dR}) is two-dimensional, while the vector $\delta{\bf R}$ is
three-dimensional. 
  Also for $\tau=\tau_{-}$, corresponding to the footpoint of the
field line on the negative polarity side, i.e.\ when
$Z(\tau_{-},{\bf r}_{+}) = z_{-}$, the 
vector $\delta{\bf R}=\delta{\bf R}_{-}$ is generally not tangent to the
photospheric plane. 
  Therefore, to find the photospheric displacement
$\delta {\bf r}_{-}$ corresponding to the negative polarity it is
necessary to project $\delta{\bf R}_{-}$ along the local direction of
magnetic field ${\bf B}_{-} \equiv {\bf B}(x_{-},y_{-},z_{-})$ onto the
photospheric plane
at the footpoint $(x_{-},y_{-},z_{-})={\bf R}(\tau_{-},{\bf r}_{+})$.
 This yields $\delta{\bf r}_{-} = {\cal P}\, \delta{\bf
R}_{-}$, where
 \begin{eqnarray}
  {\cal P} =
 \left( \begin{array}{ccc}
    1 & 0 & -\frac{B_{x-}}{B_{z-}} \\
    0 & 1 & -\frac{B_{y-}}{B_{z-}}
        \end{array}
  \right)
        \label{P}
 \end{eqnarray} 
is the projection matrix.
Applying the projection matrix to Eq. (\ref{dR}) and denoting ${\cal
D}(\tau_{-},{\bf r}_{+})$ as ${\cal D}_{-}$, we get 
$\delta{\bf r}_{-} = {\cal P}{\cal D}_{-}\, \delta{\bf r}_{+}$, which
in turn implies that the required matrix is
 \begin{eqnarray}
  \Dpm \equiv
 \left( \begin{array}{cc}
    {\partial X_{-}\over\partial x_{+}} &  {\partial X_{-}\over\partial 
y_{+}} \\
    {\partial Y_{-}\over\partial x_{+}} &  {\partial Y_{-}\over\partial 
y_{+}}
        \end{array}
  \right)
 = {\cal P} {\cal D}_{-}.
        \label{D2Dpm}
 \end{eqnarray}
Using this expression for $\Dpm$ one can find all the
characteristics  of the field line connectivity we need and, in particular,
the squashing degree~(\ref{Q}).

We now apply this theory to the calculation of the squashing degree $Q$ 
on the
quasi-separator of the ``straightened'' HFT\@.
As already mentioned in \S~\ref{s:S2}, the quasi-separator
coincides with the $z$-axis in this case.
The matrix of magnetic field derivatives on this axis can be reduced
to the form
 \begin{eqnarray}
  {\cal G}(0,0,z) =
   \left( \begin{array}{ccc}
    f(z) & 0        & 0\\
    0     & -f(-z) & 0 \\
    0     & 0       & -f(z)+f(-z)
        \end{array}
  \right),
        \label{Gqs}
 \end{eqnarray}
due to 
a
symmetry of the field (\ref{B0}).
 Here we have used the definition
 \begin{eqnarray}
  f(z) =  \left. {\partial B_{0x} \over \partial  x} \right|_{{\bf 
r}=(0,0,z)}.
        \label{f}
 \end{eqnarray}

In our case it is more convenient to determine the matrix ${\cal D}_{-}$ 
by using the coordinate $z$ rather than the parameter
$\tau$ in (\ref{Deq}) .
This simply means that instead of ${\rm d}\tau$ we have to use
${\rm d}z/B_{0z}$ in (\ref{Deq}).
The diagonal form of ${\cal G}$ on the quasi-separator then enables 
us to find ${\cal D}_{-}$ by integration of 
(\ref{Gqs}) over $z$.
This and (\ref{P})--(\ref{D2Dpm}), where $B_{x-}\equiv
B_{x}(0,0,-L)=0$ and $B_{y-}\equiv B_{y}(0,0,L)=0$ are used, 
after some transformations result in
 \begin{eqnarray}
  \Dpm_{\rm qs} \equiv \Dpm\big|_{{\bf r}_{+}=0} =
   \left( \begin{array}{cc}
    e^{\lambda} & 0  \\
    0                    & e^{-\lambda} 
        \end{array}
  \right),
        \label{Dqs} 
 \end{eqnarray}
where $\lambda$ is given by Eq. (\ref{lmd}). Then by using the
definitions of $Q$ and $\Dpm$ (see (\ref{Q}) and (\ref{D2Dpm})) we
obtain the required expression~(\ref{Qqs}).

\section{CONVECTION OF MAGNETIC FIELD BY HYPERBOLIC FLOWS}
\label{s:conv}

For a given flow ${\bf r}(t,{\bf r}_{0})$, which is related to the
velocity field ${\bf  v}(t,{\bf r})$ by Eq. (\ref{ef}), 
the general solution of the ideal induction equation
 \begin{eqnarray}
  {\partial {\bf B} \over \partial t } + \nabla \times ({\bf
    B} \times {\bf v}) = {\bf 0}
  \label{ie}
 \end{eqnarray}
can be written in the Lagrangian representation as
 \begin{eqnarray}
  {\bf B}(t,{\bf r}_{0}) = {\bf B}_{0}({\bf r}_{0})\cdot
    \nabla_{{\bf r}_{0}} {\bf r}/\det(\nabla_{{\bf  r}_{0}} {\bf r} ).
    \label{BL}
 \end{eqnarray}
For our purposes, however, the Eulerian representation is
more convenient.
One can switch to this representation by substituting
${\bf r}_{0} \rightarrow {\bf r}_{0}(t,{\bf r})$ and
$ \nabla_{{\bf r}_{0}} {\bf r} \rightarrow \left( \nabla_{\bf r}
{\bf r}_{0} \right)^{-1} $ into (\ref{BL}).
Then after some transformations, using the incompressibility condition
$\det(\nabla_{{\bf  r}_{0}} {\bf r} )=1$ for the flow (\ref{vcor}), we
obtain
 \begin{eqnarray}
  B_{x} & = &  B_{0x} {\partial \bar{y}_{0} \over \partial  \bar{y}}
  -B_{0y}{\partial \bar{x}_{0} \over \partial \bar{y} } 
  + B_{0z} {l_{\rm sh} \over L}
  \left( {\partial \bar{x}_{0} \over \partial \bar{y} } 
         {\partial \bar{y}_{0} \over \partial \bar{z}}
  - {\partial \bar{y}_{0} \over \partial \bar{y} }
    {\partial \bar{x}_{0} \over \partial \bar{z} } \right),
  \label{Bx} \\
  B_{y} & = &  B_{0y} {\partial \bar{x}_{0} \over \partial  \bar{x}}
  -B_{0x}{\partial \bar{y}_{0} \over \partial \bar{x} } 
  + B_{0z} {l_{\rm sh} \over L}
  \left( {\partial \bar{y}_{0} \over \partial \bar{x} } 
         {\partial \bar{x}_{0} \over \partial \bar{z}}
  - {\partial \bar{x}_{0} \over \partial \bar{x} }
    {\partial \bar{y}_{0} \over \partial \bar{z} } \right),
  \label{By} \\
  B_{z} & = & B_{0z},
  \label{Bz}
 \end{eqnarray}
where $\bar{z} = z/L$, while the  $x,y$-coordinates are normalized by
$l_{\rm sh}$.
Thus, the magnetic component $B_{z}$ is simply transported by the flow
as $B_{z}(t,{\bf r}) = B_{0z}({\bf r}_{0}(t,{\bf r}))$.
Since the initial component $B_{0z}({\bf r}_{0})$ is almost
homogeneous in the middle of the HFT,  $B_{z}(t,{\bf r})$ will remain
nearly constant there for a sufficiently long time. 

On the other hand, the components $B_{x}$ and $B_{y}$ will change
significantly due to the large gradients of the Lagrangian coordinates.
In the plane $z=0$ the corresponding derivatives with respect to $\bar{x}$
and $\bar{y}$ can be found directly from (\ref{X0z0})
and~(\ref{Y0z0}).
The calculation of the derivatives with respect to $z$ is a less trivial
problem and will be considered below.

From the group properties of the flow (see
\S~\ref{s:km}) 
it follows that the mapping 
${\bf r}_{0}(t,{\bf r})$ is described by
 \begin{eqnarray}
  {{\rm d}{\bf r}_{0} \over {\rm d}t} = -{\bf v}({\bf r}_{0}),
  \label{ef0}
 \end{eqnarray} 
where the dependence on ${\bf r}$ is parametric and due to the
initial condition ${\bf  r}_{0}(0,{\bf r}) = {\bf r}$.
Since $v_{z}\equiv 0$ 
the plasma motion is planar
such that $z_{0}(t,{\bf r}) \equiv z$.
 Expressions for the other Lagrangian coordinates can be derived
explicitly
from (\ref{ef0}) and (\ref{vcor}) only in the plane $z=0$ and only in the
case of ``twisting''  shearing motions, giving~(\ref{X0z0})--(\ref{Y0z0}).
 Differentiating (\ref{ef0}) with respect to $z$ and restricting the
result for this plane, one obtains
 \begin{eqnarray}
{{\rm d}\ \over{\rm d} \bar{t}} {\partial \bar{x}_{0} \over \partial
  \bar{z}} 
  & = & -{1 \over (\cosh\bar{y}_{0})^{2}}
   {\partial \bar{y}_{0} \over \partial \bar{z}} + \tanh\bar{y}_{0},
  \label{eX0z} \\
{{\rm d}\ \over{\rm d} \bar{t}} {\partial \bar{y}_{0} \over \partial
  \bar{z}} 
  & = & -{1 \over (\cosh\bar{x}_{0})^{2}}
   {\partial \bar{x}_{0} \over \partial \bar{z}} - \tanh\bar{x}_{0},
  \label{eY0z}
 \end{eqnarray}
which is a system of ordinary differential equations whose unknowns
must satisfy the initial conditions
 \begin{eqnarray}
  {\partial \bar{x}_{0} \over \partial \bar{z}}={\partial \bar{y}_{0} \over
 \partial \bar{z}}=0.
  \label{ics}
 \end{eqnarray}
 In combination with (\ref{X0z0})--(\ref{Y0z0}) this system
of differential equations
allows us to calculate the magnetic components (\ref{Bx}) and (\ref{By}) 
in the plane~$z=0$ at
any time.

Due to a symmetry of the
flow, we can make further analytical progress for the particular streamline
$\bar{y}=-\bar{x}$ in the plane $z=0$ .
 The system of differential equations can in this case be integrated 
 explicitly to yield
 \begin{eqnarray}
   {\partial \bar{x}_{0} \over \partial \bar{z}}=
   {\partial \bar{y}_{0} \over \partial \bar{z}} = 
   { \left[ 1 + (e^{\bar{t}}\sinh\bar{y})^{2} \right]^{1/2} \over
    2 e^{\bar{t}} \sinh\bar{y} }\,
   \ln \left({ 1 + (e^{\bar{t}}\sinh\bar{y})^{2} 
   \over 1 + \sinh^{2}\bar{y} } \right),
  \label{drvz}
 \end{eqnarray} 
which enables us to find $B_{x}$ and $B_{y}$ according to the
algorithm described above.
Using the simplified form of magnetic field (\ref{B0a}) and the
 system of $(x^{\prime},
y^{\prime})$-coordinates rotated by $45^{\circ}$ with respect
to $x$ and $y$, one obtains the required field
component~(\ref{Bxp}).
 The current density along the quasi-separator
(\ref{jzqs})--(\ref{jzqs2}) is found in a similar way.


\clearpage

\clearpage

\clearpage

\clearpage

\clearpage 

\clearpage 

\clearpage

\clearpage

\clearpage

\clearpage

\clearpage

\clearpage

\end{document}